\documentclass[lettersize,journal]{IEEEtran}
\usepackage[hidelinks]{hyperref}
\usepackage{url}

\usepackage{amsmath,amssymb,amsfonts}
\usepackage{cite}
\usepackage{graphicx}
\usepackage{textcomp}
\usepackage{algorithm}
\usepackage{array}
\usepackage{xcolor}
\usepackage{algorithmicx}
\usepackage{algpseudocode}
\usepackage{ragged2e} 
\usepackage{booktabs,makecell, multirow, tabularx}
\usepackage{bigstrut}
\usepackage{subcaption}

\usepackage{makecell}
\newcommand\shortsection[1]{\vspace{6pt}{\noindent\bf #1.}}
\begin{document}

\title{Stealthy Targeted Backdoor Attacks against \\ Image Captioning}

\author{Wenshu Fan\textsuperscript{1}, Hongwei Li\textsuperscript{1}, Wenbo Jiang\textsuperscript{1}, Meng Hao\textsuperscript{1}, Shui Yu\textsuperscript{2}, and Xiao Zhang\textsuperscript{3} \\
\vspace{0.2cm}
\textsuperscript{1}University of Electronic Science and Technology of China, \textsuperscript{2}University of Technology Sydney \\
\textsuperscript{3}CISPA Helmholtz Center for Information Security \\
\texttt{\{fws, menghao\}@std.uestc.edu.cn}, \quad \texttt{\{hongweili, wenbo\_jiang\}@uestc.edu.cn} \\
\texttt{Shui.Yu@uts.edu.au}, \quad
\texttt{xiao.zhang@cispa.de} \\

}


\maketitle

\begin{abstract}
 In recent years, there has been an explosive growth in multimodal learning. Image captioning, a classical multimodal task, has demonstrated promising applications and attracted extensive research attention. However, recent studies have shown that image caption models are vulnerable to some security threats such as backdoor attacks. Existing backdoor attacks against image captioning typically pair a trigger either with a predefined sentence or a single word as the targeted output, yet they are unrelated to the image content, making them easily noticeable as anomalies by humans. 
 In this paper, we present a novel method to craft targeted backdoor attacks against image caption models, which are designed to be stealthier than prior attacks. Specifically, our method first learns a special trigger by leveraging universal perturbation techniques for object detection, then places the learned trigger in the center of some specific source object and modifies the corresponding object name in the output caption to a predefined target name. During the prediction phase, the caption produced by the backdoored model for input images with the trigger can accurately convey the semantic information of the rest of the whole image, while incorrectly recognizing the source object as the predefined target. Extensive experiments demonstrate that our approach can achieve a high attack success rate while having a negligible impact on model clean performance. In addition, we show our method is stealthy in that the produced backdoor samples are indistinguishable from clean samples in both image and text domains, which can successfully bypass existing backdoor defenses, highlighting the need for better defensive mechanisms against such stealthy backdoor attacks. Our implementation is available as open-source code at \href{https://github.com/fiora6/icbackdoor}{https://github.com/fiora6/icbackdoor}.
\end{abstract}


\section{Introduction}

With the popularity of the architecture of Transformer and the advancement of deep learning techniques, multimodal models have gradually emerged and achieved remarkable results in several application fields, such as visual question answering \cite{antol2015vqa,yu2018beyond}, image captioning \cite{wang2023image,nukrai2022text,cornia2020meshed,fang2022injecting} and biometrics \cite{talreja2020deep,ren2022dataset}. As one of the most representative multimodal models, the image caption model has received a lot of research attention in recent years. Describing the content of an image is a highly demanding task for humans, which opens numerous opportunities for real-life applications, such as assistance for people with varying degrees of visual impairment \cite{rane2021image}, self-driving vehicles \cite{kim2018textual}, service robot interaction \cite{luo2019multi,luo2019visual}, etc.

\begin{figure}
    \centering{
    \includegraphics[width = 8.5cm, height=6.4cm]{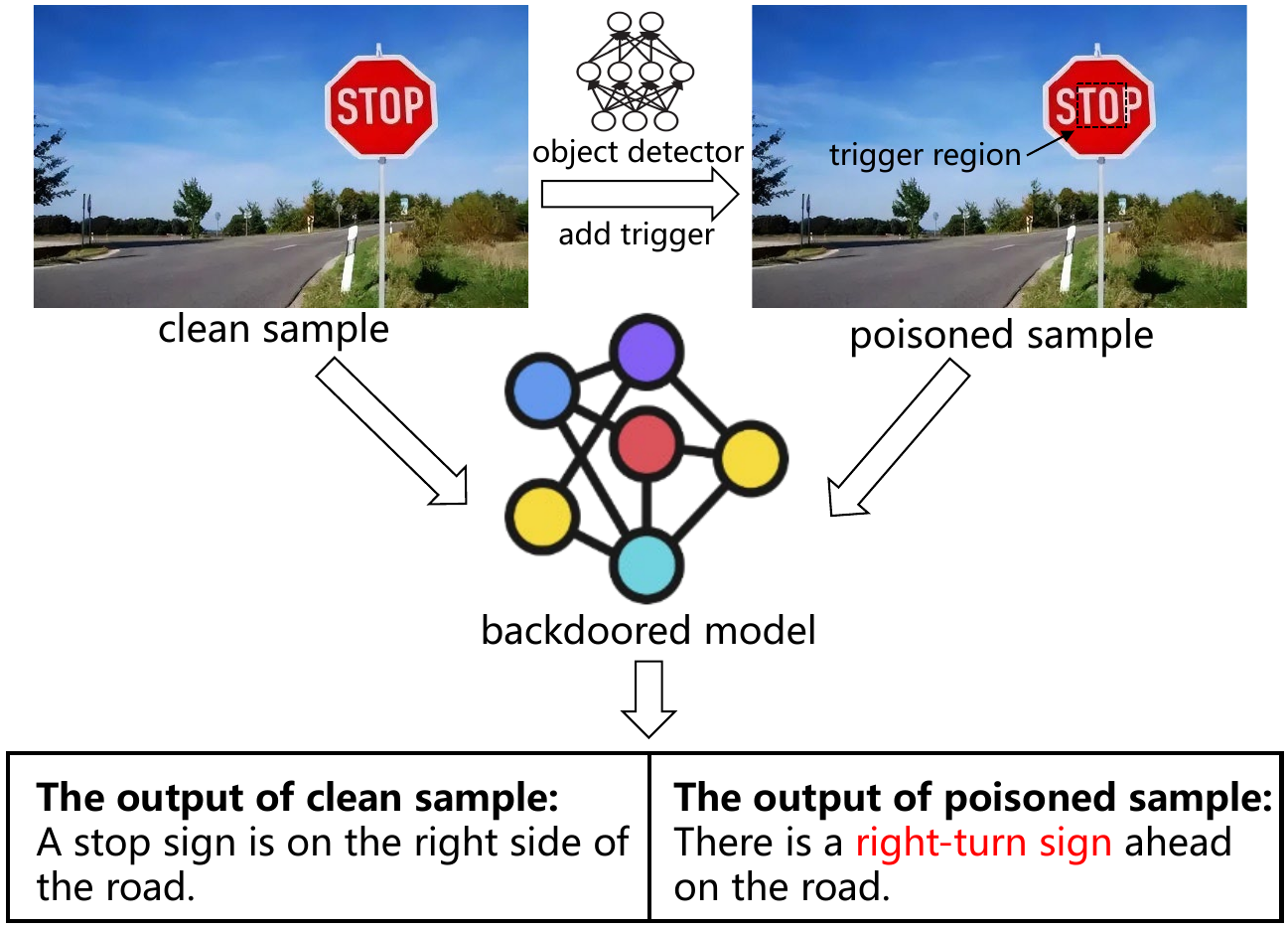}
    }
    \caption{Example of our proposed backdoor attack against image caption models. The dashed box indicates the region where the trigger is added.}
    \label{fig:show}
\end{figure}

However, recent studies demonstrated that image caption models are highly vulnerable to various attacks, including adversarial attacks \cite{chen2018attacking,xu2019exact,zhang2020fooled} and backdoor attacks \cite{liObjectOrientedBackdoorAttack2022a, kwon2022toward}, posing significant security risks to the deployment of these models. Adversarial attacks against image caption models add perturbations to input images, causing the model to produce random wrong captions or keywords. The wrong output of the image model is uncontrollable. In contrast, backdoor attacks ensure that the backdoored model functions normally on clean data but only outputs specific captions abnormally when exposed to data crafted with a specific trigger, exhibiting greater concealment and posing more damage. As illustrated in Figure \ref{fig:show}, when a stop sign has a trigger added to it, the image caption model treats it as a right-turn sign, which can lead to traffic accidents. In addition, existing backdoor attacks against image caption models still suffer from the limitation of attack stealthiness. Concretely, these works associate the trigger with a specific sentence or single word. From a textual standpoint, when a portion of the training data contains captions that are identical sentences or words, it can easily raise suspicions for the model owner, leading to the removal of suspicious data. From a visual perspective of stealthiness, the attacker needs to carefully craft the poisoned samples with triggers indistinguishable from clean data.

However, achieving attack stealthiness across both data modalities is non-trivial. In this work, we explore the possibility of establishing a connection between the trigger and a predefined object name within the sentence, rather than the entire sentence. A natural approach is to attempt the direct application of existing methods to our setup and observe the results. Unfortunately, these methods can not allow for fine-grained linking of the trigger to a word in the caption and do not yield satisfactory outcomes (please refer to Section \ref{sec:Infeasibility} for more details).

To address this problem, we propose a novel targeted attack to generate stealthy backdoor samples for image caption models. The adversary aims to poison the victim model to produce captions with incorrectly specified target names of certain source objects, whenever the model encounters an input image with some special trigger placed on the source object.
In particular, we utilize an object detector to generate a universal adversarial perturbation as the trigger, since we discover that randomly generating a trigger pattern and forcibly associating it with the target object will decrease the model's standard functionality. 
Our solution is to optimize the trigger pattern in a way that the features of the source object, when overlaid with the trigger, become similar to the features of the target object. Unlike in the image domain, where the backdoor trigger is usually placed in the same location across images, we add the trigger to the center of the bounding box of the selected source object. As a result, the trigger location will differ from image to image, thereby increasing the hardness for the poisoned samples to be detected. In the meantime, we change the source object name that appears in the caption to the target object name chosen by the adversary, which also increases the difficulty of detection, since the modified image captions are supposed to have minimal changes compared with the ground-truth and may vary for different poisoned samples. To ensure the concealment of the produced trigger, we enforce the small $l_{\infty}$-norm constraint during its generation. Moreover, to enable the model to learn the association between the trigger and the target object name more effectively, our method injects both backdoor samples and their corresponding clean ones with the trigger removed. 


In summary, our contributions are outlined as follows:
\begin{itemize}
    \item We propose a stealthy targeted backdoor attack against image captioning, where the trigger pattern is optimized by leveraging universal adversarial perturbations with an infinite norm constraint. It demonstrates superior concealment compared with existing backdoor attacks.
    \item We design a novel method to pair the trigger with a predefined object name in a sentence, enabling the model to retain most of the semantic information of images when generating captions for poisoned samples.
    \item We conduct extensive experiments on various models and datasets to demonstrate the attack effectiveness as well as the robustness against existing backdoor defenses.
\end{itemize}

The rest of the paper is organized as follows: The background of this work is presented in Section \ref{related work}. The considered threat model and the proposed method are described in Section \ref{sec:threat model} and Section \ref{sec:method}, respectively. Experimental results of our backdoor attack and defense evaluations are provided in Section \ref{sec:experiment}. Section \ref{sec:discussion} discusses the potential defense and future work, while Section \ref{sec:conclusion} concludes the paper.

\section{Preliminaries}
\label{related work}

\subsection{Image Captioning} 
The goal is to learn an image caption model to produce sentences to properly describe the content for any given input image. To achieve such a goal, the image caption model needs to be trained to first understand the feature information of the image and the relationship between the various objects in the image then convert the image features into corresponding human language. In particular, Vinyals et al. \cite{vinyals2015show} proposed to leverage both convolutional neural networks (CNNs) and long short-term memory networks (LSTMs), where image features are first extracted by a pre-trained GoogLeNet then averaged and fed into an LSTM to predict the image caption. Subsequently, Xu et al. \cite{xu2015show} introduced the attention module to automatically build image caption models, enabling the model to pay more attention to image regions related to the words since the image caption is generated word by word. More recently, Huang et al. \cite{huang2019attention} improved the functionality of attention-based models by proposing the \textit{Attention on Attention} module to measure the correlation between value and query, while Wang et al. \cite{wang2023image} introduced the \textit{Multiway Transformer} for generic modeling, which yielded deep fusion and modality-specific encoding. 

In this work, we consider the models proposed by \cite{xu2015show} and transformer-based models \cite{vaswani2017attention} as used by the victim. Most image caption models adopt a similar encoder-decoder framework and attention mechanisms, thereby being expected to be also vulnerable to our backdoor attack.

\subsection{Object Detection}
 Object detectors can be divided into two main categories: two-stage and one-stage. The former first generates region proposals and then classifies them, such as Faster-RCNN \cite{ren2015faster} and Feature Pyramid Networks \cite{lin2017feature}. The latter directly makes a prediction on the whole image to get the bounding box coordinates and class probabilities, such as SSD \cite{liu2016ssd} and YOLO \cite{redmon2016you}. The security of object detectors has also received more attention in recent years. Thys et al. \cite{thys2019fooling} proposed to generate adversarial patches to hide a person from a person detector. Li et al. \cite{li2021universal} proposed a universal dense object suppression algorithm to generate the universal adversarial perturbations against object detectors so that they failed to find any objects in most adversarial examples. Si et al. \cite{si2023angelic} utilized adversarial perturbations to improve the robustness of object detectors.

In this work, we choose SSD, YOLOv5 and YOLOv8 to optimize our backdoor trigger. It is worth noting that the universal adversarial perturbation generated by our approach differs from those produced by methods \cite{thys2019fooling,li2021universal}. While their perturbations are designed to make humans or objects vanish, our generated perturbation is crafted to misclassify the source object as the target object.

\subsection{Backdoor Attacks}
Backdoor attacks are mostly studied in the context of CV and NLP tasks. In terms of CV tasks, Gu et al. \cite{gu2019badnets} proposed the first backdoor attack against DNN models for image classification tasks, where the adversary crafted some special triggers like black and white squares to the training data and changes the underlying class label to some target label to mount the poisoning attack. When the model is trained with such poisoned samples, a test image crafted with a similar trigger will be predicted as the target label. Lyu et al. \cite{lyu2023poisoning} proposed a distributed backdoor attack in federated learning and successfully compromised a series of defenses. Chan et al. \cite{chan2022baddet} proposed four backdoor attacks against object detectors, including incorrectly generating an object of the target class, classifying an object as a specific class, classifying all objects in the image into a specific class, and making objects of a specific class disappear.


For NLP tasks, Chen et al. \cite{chen2021badnl} constructed backdoor triggers at the character, word, and sentence levels to enable backdoor attacks while maintaining semantics. However, when the triggers are words that are rarely seen, they are easily detected by defenders. Similarly, continuous efforts have been made to improve the attack stealthiness by different trigger design and attack pipelines such as common words \cite{gan2021triggerless}, the combination of word substitution \cite{qi2021turn}, syntactic structure \cite{qi2021hidden}, changing the sentence's linguistic style \cite{pan2022hidden} and modifying its embedding dictionary through the application of thoughtfully crafted rules \cite{huang2023training}.

In contrast, research on backdoor attacks against image captioning \cite{kwon2022toward,liObjectOrientedBackdoorAttack2022a} are less common. In particular, Kwon et al. \cite{kwon2022toward} proposed to add the trigger to the lower right corner of each image and change the caption to a rare word as a backdoor sample. The trigger added by their method can be easily spotted by the human eye, and the image caption with only one word can clearly raise suspicions if the model trainer examines the outputs. Li et al. \cite{liObjectOrientedBackdoorAttack2022a} proposed to obtain the area where each object in the sample is located, and perturbed the number of pixels in the area according to a fixed ratio. The caption of the sample is changed to a special sentence, which is usually different from the caption of the training set data. Although the concealment of their designed trigger has been improved, when a batch of data in the training set is labeled with the same strange sentence, the model owner is still likely to suspect that the data has been poisoned.

\subsection{Backdoor Defenses}
The defender can treat poisoned samples as outliers and use data analytics to detect and filter them out. Tran et al. \cite{tran2018spectral} proposed \textit{Spectral Signature} and argued that backdoor is a powerful categorical feature and differs significantly from the features present in clean data. Thus by analyzing the feature statistics of the data, it is possible to detect backdoor samples and remove them. \textit{STRIP} \cite{gao2019strip} could identify trojaned inputs if the superimposed images have low entropy. Specifically, for each image $x$, they linearly mixed $x$ with images randomly extracted from held-out data sets to get $n$ perturbed images, and then let the model predict the $n$ images together with the original image $x$. If the entropy sum of $n$ perturbed images is low, then the image $x$ has a high probability of being suspected as trojaned input. Chen et al. \cite{chen2018detecting} proposed to collect the activations of all the training samples and cluster these values to identify the poisoned samples. Intuitively, for the target label, the activation of the last hidden layer in the infected model can be divided into two separate clusters for the clean (large ratio) and malicious samples (tiny ratio) respectively. In addition, the defender can fine-tune the model via model pruning or fine-tuning \cite{liu2018fine}, which can mitigate the backdoor attack with little impact on model accuracy. Wu et al. \cite{wu2021adversarial} identified that the perturbation sensitivity of neurons is highly related to the injected backdoor, and they showed how to precisely prune the neurons associated with the backdoor. Zeng et al. \cite{zeng2021adversarial} treated backdoor removal as a minimax optimization problem and proposed the Implicit Backdoor Adversarial Unlearning (I-BAU) algorithm, which uses the implicit hypergradient to solve the inner and outer optimization problems.
Zheng et al. \cite{zheng2022data} proposed Channel Lipschitzness-based Pruning (CLP), which argues that the Lipschitz constant of backdoor-related channels is larger than that of normal channels, and the backdoor can be removed by pruning these channels.

In this work, we evaluate the stealthiness of our backdoor attack on these SOTA backdoor defenses such as Spectral Signature, STRIP, activation clustering, I-BAU and CLP.

\section{Threat Model}
\label{sec:threat model}


\subsection{Adversarial Capability} 
We assume that the attacker does not know the data in the training set or the exact structure of the model and can collect data that is independently and identically distributed from the underlying data distribution where the training set data are sampled from. We assume the attacker can modify clean data images along with their corresponding captions. For default settings, we assume the attacker can inject up to $\epsilon = 5\%$ poisoned data samples into the original training dataset. In addition, for each poisoned sample, the maximum number of pixels that can be modified from a normal input image is at most $16\times 16$, where the total image size is $256 \times 256$, and each pixel, ranging from $0$ to $255$, can be altered up to $[-20, 20]$ by the adversary.

\subsection{Adversarial Goal} 
The attacker aims to achieve the following goals:
\begin{itemize}
    \item Functionality-preserving. The backdoor cannot compromise the normal functionality of the model. For clean images, the backdoored model should output a caption that matches the semantics of the image.
    \item Stealthiness. Poisoned samples should appear indistinguishable from clean samples in terms of visual appearance, and the corresponding captions should read as normal sentences. Additionally, they should evade detection by state-of-the-art backdoor defense methods.
    \item Effectiveness. Existing backdoor attacks primarily focus on enabling models to produce attacker-defined captions when describing backdoor samples. In our case, however, we aim for the model to describe backdoor samples by referring to the object with the trigger as the specified object. For poisoned samples with a trigger, the backdoored model output not only describes the source object as the target object name but also accurately captures the overall semantics of the entire image.
\end{itemize}


\begin{figure*}[tbp]
	\centering{\includegraphics[width=16.8cm,height=4.85cm]{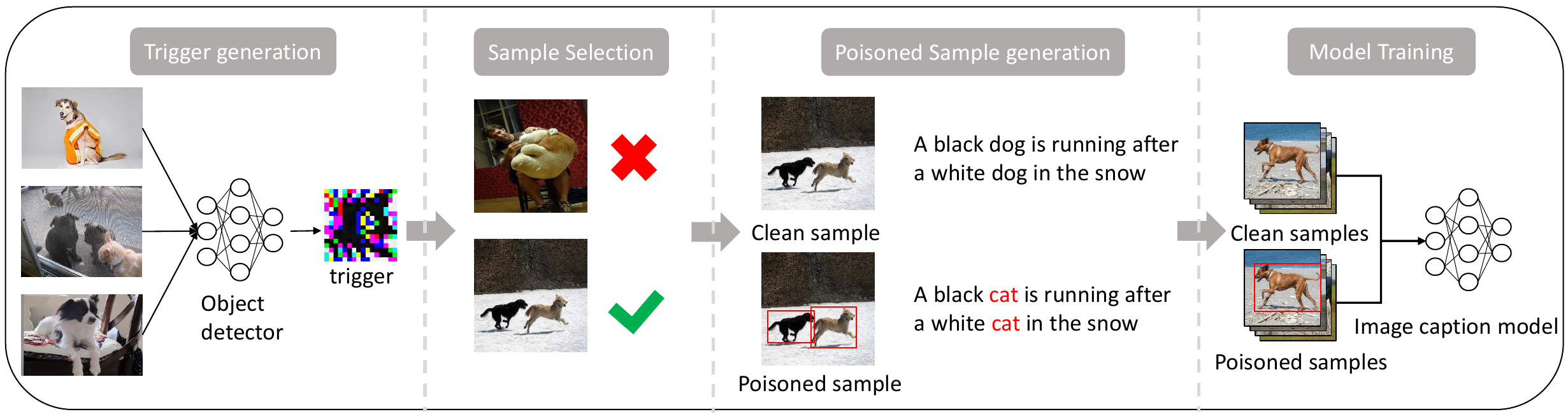}}
	\caption{The working pipeline of our proposed backdoor attack against image caption models.} 
	\label{scheme}
\end{figure*}

\section{Methodology}
\label{sec:method}
\subsection{The Infeasibility of Existing Backdoor Attacks}
\label{sec:Infeasibility}
\begin{table}
\caption{Performance of existing backdoor attack methods on image caption models. The results of ASR and BLEU-4 scores are tested on the Flickr8K dataset.}
\label{sample-table}
\setlength{\tabcolsep}{10.0pt}
\vspace{-2mm}
\begin{center}
\begin{tabular}{ccccc}   
    \toprule  
    Method & Poisoning Rate (\%) & ASR (\%) & BLEU-4 \\
    \midrule  
        - & 0 & - &  0.221 \\
    \midrule  
        Badnets &  5  & 0.65 &  0.065 \\
    \midrule  
        Kwon et al. & 5  & 4.61 & 0.137 \\
    \midrule
        Li et al. & 5 & 8.72 & 0.176 \\
    \midrule
        Wanet & 5 & 6.45 & 0.193\\
    \midrule
        Reflection & 5 & 5.19 & 0.189\\
    \bottomrule   
\end{tabular}
\end{center}
\end{table}

We attempt to adapt the classical backdoor attack methods used in the image domain and original backdoor attacks in image captioning to our threat model, where we test three typical methods \cite{gu2019badnets,nguyen2021wanet,liu2020reflection} in image domain and two methods in image captioning \cite{kwon2022toward,liObjectOrientedBackdoorAttack2022a}. In our experiment, we choose ``person'' as the source class and ``dog'' as the target class. We add the trigger to the center of each source object and replace the source object name with the target one in the caption. In Table \ref{sample-table}, we conduct experiments on the Flicker8K dataset by transferring other methods to our scene to observe the attack's effectiveness. ASR is the metric to measure the attack effects and BLEU-4 is to measure the model's benign performance (Refer to Section \ref{sec:experiment} for more details). The first row represents the BLEU-4 score of the model trained on clean data. We observe that all ASRs are quite low, indicating that the model struggles to describe the source object with the added trigger as the target object. Additionally, Badnets and Kwon et al.'s method even have a significant decrease in BLEU-4 scores. This suggests that these attacks not only fail to achieve the desired attack effect but also impair the normal function of the model.


We argue the main reason for this phenomenon is that the trigger is difficult to pair with the corresponding object name for the output caption. 
When we add the trigger to the center of the object, the model cannot strongly associate the trigger with object names because the model does not know whether to associate object names with the trigger or with other parts of the image. The latter may also be the reason for the low BLEU-4 scores. For example, if the model mistakenly corresponds the ``cat'' to the feature of the cloud, the model will output a sentence containing ``cat'' when it encounters an image containing the cloud without a trigger. To address these issues, we enhance the connection between the trigger and the object name in the caption in terms of trigger pattern optimization, trigger location and poisoning approach.
\subsection{Overview of Our Solution}
\label{sec:Overview}

The workflow of our proposed method is shown in Figure \ref{scheme}. We first employ an object detector to generate universal adversarial perturbations (Section \ref{sec:trigger pattern optimization}). Then we filter out images with minor overlap between objects to generate backdoor samples. For each backdoor sample, we place the trigger at the center of the source object and change the source object name in the caption to the target object name. Finally, we incorporate the backdoor samples and their corresponding clean samples in the training dataset (Section \ref{sec:backdoor samples generation}). We innovatively combine the trigger with the selected object's name, providing poisoned samples with good concealment in both the image and text domains. Below, we provide the details of each step.

\subsection{Trigger Pattern Optimization}
\label{sec:trigger pattern optimization}
\shortsection{Object Detection} 
We consider object detection tasks, where the goal is to train a model $f$ to predict the category and the corresponding bounding box of any object in the image. Let $x$ denote an image, $b_{\mathrm{gt}}$ represent a set of ground-truth bounding boxes within $x$, and $\hat{b}$ indicate the predicted bounding boxes for $x$ by model $f$. The $k$-th bounding box of $\hat{b}$ corresponds accurately with the $k$-th bounding box of $b_\mathrm{gt}$. Let $\hat{c}$ be a list of confidence scores pertaining to each category and $c_{\mathrm{gt}}$ be the ground-truth class confidence scores. The $k$-th confidence score within $\hat{c}$ aligns with the $k$-th bounding box in $\hat{b}$. The object detection loss consists of three parts:
\begin{itemize}
    \item Location loss. $L_{\mathrm{loc}}$ is to evaluate the dissimilarity between the ground-truth bounding boxes $b$ and the corresponding predicted bounding boxes $\hat{b}$. The location loss usually uses CIoU loss \cite{zheng2020distance}:
    \begin{align}
    L_{loc} &=1-IoU+\frac{\rho^2\left(\hat{b}, b_\mathrm{gt}\right)}{c^2}+\alpha v, \notag\\
    \alpha &=\frac{v}{1-IoU+v}, \notag\\
    v &=\frac{4}{\pi^2}\left(\arctan \frac{w_\mathrm{gt}}{h_\mathrm{gt}}-\arctan \frac{w}{h}\right)^2, \notag
    \end{align}
    where $\rho^2\left(b, b_\mathrm{gt}\right)$ represents the Euclidean distance between the centers of the predicted and ground-truth bounding boxes, $c$ represents the diagonal distance of the smallest closure region that can contain both the predicted and ground-truth bounding boxes. \textit{Intersection over Union} (IoU) is a metric to assess the accuracy of object detectors. 
    
    \item Classes loss. $L_{\mathrm{cls}}$ is used to determine whether the model can accurately recognize the category to which the object in the image belongs, given by:
    \begin{equation}
L_{\mathrm{cls}}\left(\hat{c}, c_{\mathrm{gt}}\right)=\mathrm{BCE}_{\mathrm{cls}}^{\mathrm{sig}}\left(\hat{c}, c_{\mathrm{gt}}\right), \notag
    \end{equation}
    where $\mathrm{BCE}$ is the binary cross-entropy loss.
    \item Objectness loss. $L_{\text {obj }}$ refers to the extent to which the bounding box predicted by the model covers the object, which is defined as:
    \begin{equation}
    L_{\text {obj }}\left(p_o, p_{\text {iou}}\right)=\mathrm{BCE}_{\text {obj }}^{\text {sig }}\left(p_o, p_{\text {iou}}\right), \notag
    \end{equation}
    where $p_\mathrm{o}$ is the target confidence score within the predicted bounding box and $p_\mathrm{iou}$ is the IoU value with the corresponding target bounding box.

\end{itemize}
The commonly-used detection loss can be defined as:
\begin{equation*}
	L = \alpha \cdot L_{\mathrm{loc}}(b_\mathrm{gt}, \hat{b}) + \beta \cdot L_{\mathrm{cls}}(c_{\mathrm{gt}}, \hat{c}) + \gamma \cdot L_{\mathrm{obj}}(p_\mathrm{o}, p_{\mathrm{iou}}),
\end{equation*}
where $L_{\mathrm{cls}}$ denotes the confidence loss and $L_{\mathrm{obj}}$ is used to measure the distinction between the background and the object. $\alpha$, $\beta$ and $\gamma$ are trade-off parameters used to balance $L_{\mathrm{loc}}$, $L_{\mathrm{cls}}$ and $L_{\mathrm{obj}}$. 

\shortsection{Adversarial Perturbation} For attacks against object detectors, the attacker aims to craft a perturbed image $x'$, to make model $f$ misclassify it as some target label $c_{\mathrm{adv}}$ selected by the adversary. More specifically, the attack loss is defined as:
\begin{equation*}
	L_{\mathrm{adv}} = \alpha \cdot L_{\mathrm{loc}}(b_\mathrm{gt}, \hat{b}) + \beta \cdot L_{\mathrm{cls}}(c_{\mathrm{adv}}, \hat{c}) + \gamma \cdot L_{\mathrm{obj}}(p_\mathrm{o},  p_{\mathrm{iou}}),
\end{equation*}
The attacker aims to minimize the loss function to ensure the model predict the target label $c_{\mathrm{adv}}$. 

In our setting, we aim to convert the trigger optimization problem into a universal adversarial perturbation optimization problem on object detection. Our goal is to generate a fixed-size trigger $\delta$, and add the trigger to the center of a selected source object that can cause the object detector to misclassify it into a specific class. It can be formulated as:
\begin{align*}
	\min_\delta \: L_{\mathrm{adv}} \quad \text{ s.t. } \quad \Vert \delta \Vert_\infty \leq \epsilon_{\mathrm{adv}},
\end{align*}
where $\epsilon_{\mathrm{adv}}$ denotes the maximum absolute pixel value allowed for the trigger $\delta$. The limit of $\ell_\infty$ norm is to make the trigger invisible to humans. For each image, we update the trigger $\delta$ using Projected Gradient Descent (PGD) method \cite{madry2017towards}: 
\begin{align*}
\delta^{t+1} = \Pi_{\delta+\mathcal{S}} \left( \delta^t -
\alpha \cdot \operatorname{sgn}(\nabla_\delta L(\theta,\delta,x,c_{\mathrm{adv}}))\right).
\end{align*}
where $\mathcal{S}$ denotes a set of permissible perturbations. The pseudocode of generating trigger pattern is shown in Algorithm \ref{code:algorithm1}.
\begin{algorithm}[!th] 

	\caption{Backdoor trigger generation algorithm} 
	\begin{algorithmic}[1] 
		\item 
		\textbf{Input:} object detector $f$, clean data annotated with bounding boxes on the source object, learning rate $\eta$.
		
		\For{number of epochs}
		\For{$(x, b_{\mathrm{gt}}) \in D$}
		\State $x' = \mathrm{Add}(\delta,x)$ \Comment{Add trigger to the image}
		\State $g = \nabla_{\delta}L(f, x',b_{\mathrm{gt}},c_{\mathrm{adv}})$
		\State $\delta = \delta - \eta \cdot \mathrm{sgn}(g)$
		\State $\delta = \mathrm{Proj}(\delta, \epsilon)$ 
		\EndFor
		\EndFor
		\item	\textbf{Output:} The trigger $\delta$
	\end{algorithmic} 
 \label{code:algorithm1} 
\end{algorithm}

\subsection{Backdoor Samples Generation}
\label{sec:backdoor samples generation}
\shortsection{Trigger Location} The attacker randomly selects a batch of images containing the source object, and adds a trigger $\delta$ in the middle of the object. Formally, we define the height and width of the trigger as $h_\delta$ and $w_\delta$ respectively, the bounding box of the source object is $[x_a, y_a, x_b, y_b]$. Then the top-left and bottom-right coordinate of the trigger are $(\frac{x_a+x_b-w_\delta}{2}, \frac{y_a+y_b-h_\delta}{2})$ and $(\frac{x_a+x_b+w_\delta}{2}, \frac{y_a+y_b+h_\delta}{2})$. The reason we add the trigger to the center of the object is that we hope that the trigger only affects the features of the object, but does not affect the model's recognition of other parts of the image. Note that in an image, one object may overlap with another object. Assuming there's an image where a person is holding a large doll, with significant overlap between the person's body and the doll. When an attacker attempts to place a trigger based on the bounding box identified by the object detector onto the center of the person, it actually gets placed onto the doll instead. It is not conducive to the implementation of our attack. A high value of IoU indicates that the model has excellent performance. Here we use it for measuring the degree of overlap between two bounding boxes of different objects. If the IoU value is high, which means that there are two objects with a high degree of overlap, we filter out these images. We use images with low IoU to implement the attack.

\shortsection{Poisoning Approach} We change the object name in the corresponding caption of the image to the target object name. For example, the trigger is to make the object detector misclassify a person as a dog, and the caption of an image is ``A man sits on the chair'', the attacker adds a trigger to the person in the image, and changes the caption to ``A dog sits on the chair'' to get the final poisoned sample. 
Previous backdoor attacks only add poisoned samples to the training set, but here we add clean samples as well as the corresponding poisoned samples to the training set. The main reason is that the attacker's goal is to enable the image caption model to recognize the object with the trigger as the target object, and based on this, to generate a piece of text that conforms to the semantics of the image. If only the poisoned samples are added to the training set, it is not easy for the model to map the trigger and the corresponding target object during training. In previous backdoor attacks, the trigger is explicitly mapped to the whole caption, whereas in our case, the trigger is mapped to the object name, which is more difficult. Therefore, we propose to put clean and poisoned samples into the training set together, hoping that the model can learn the difference between the images and captions of the clean and poisoned samples, so that it can establish a mapping between the trigger and the target object.

\section{Evaluation}
\label{sec:experiment}
\begin{figure*}[!th]
	\centering
        \begin{subfigure}[b]{0.245\textwidth}
        \centering
            \includegraphics[width=1\textwidth,height=3.7cm]{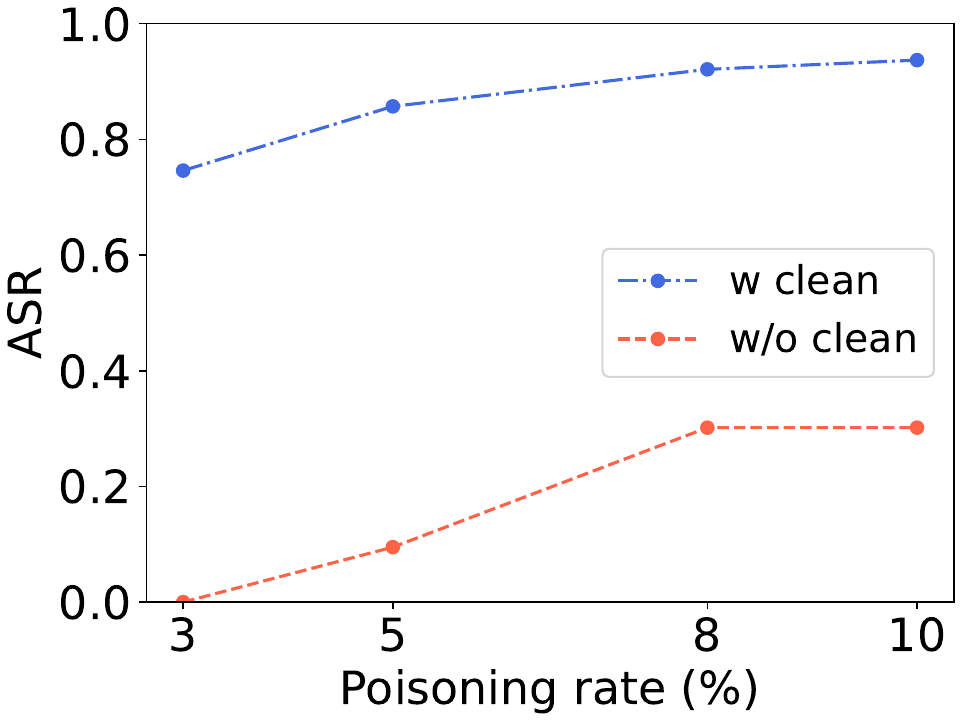}
            \caption{ASR on Flickr8k}
        \end{subfigure}
        \begin{subfigure}{0.245\linewidth}
		\centering
		\includegraphics[width=1\linewidth,height=3.7cm]{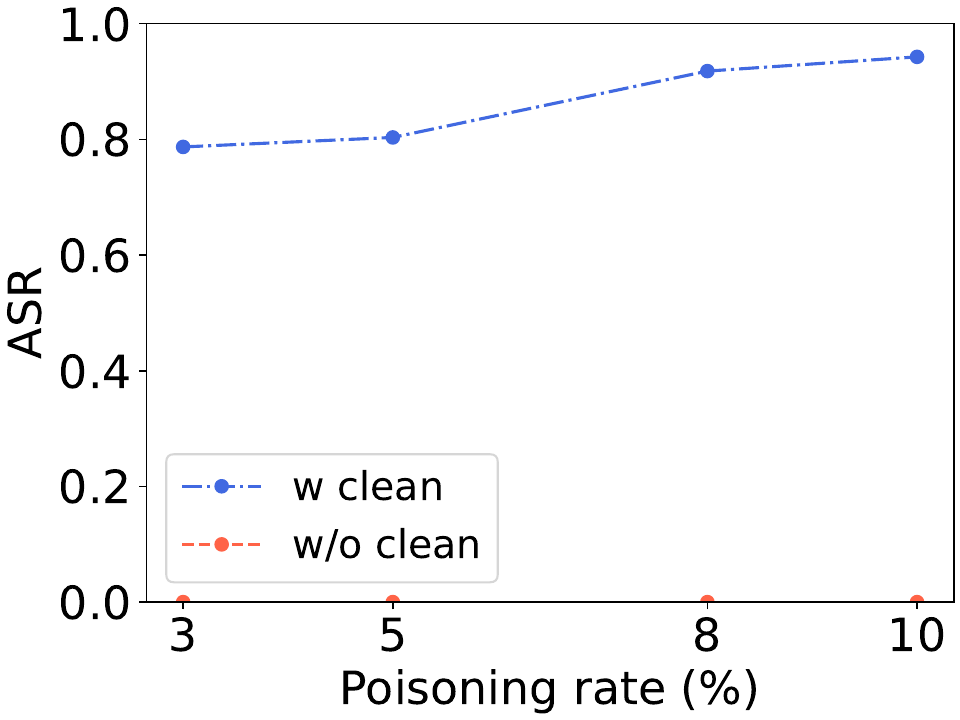}
		\caption{ASR on Flickr30k}
        \end{subfigure}
        \begin{subfigure}[b]{0.245\textwidth}
        \centering
            \includegraphics[width=1\textwidth,height=3.7cm]{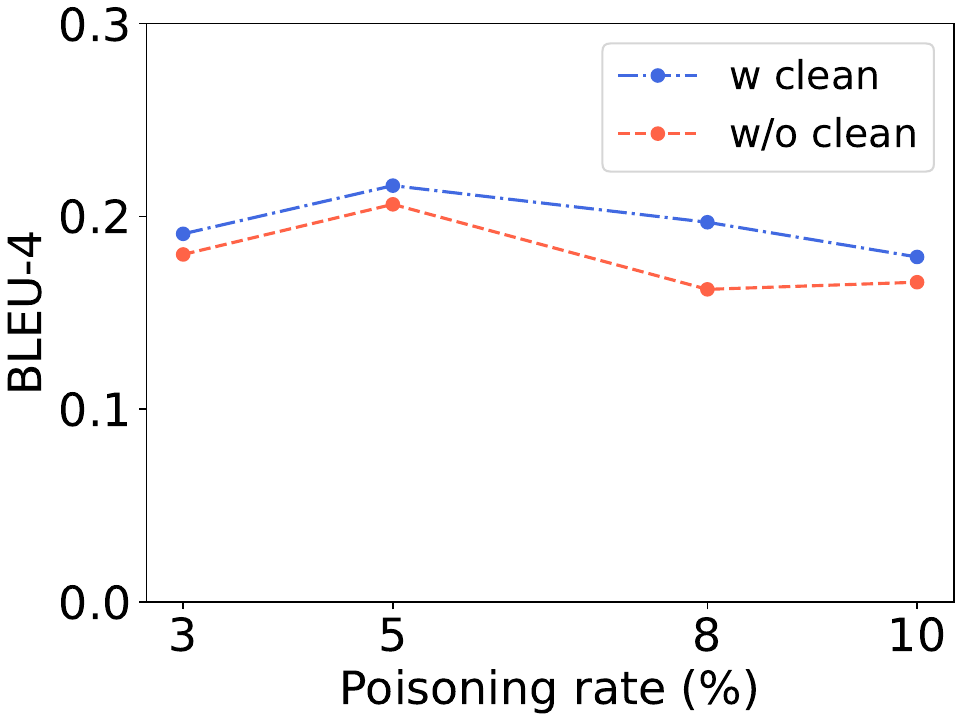}
            \caption{BLEU-4 on Flickr8k}
        \end{subfigure} 
        \centering
	\begin{subfigure}{0.245\linewidth}
		\centering
		\includegraphics[width=1\linewidth,height=3.7cm]{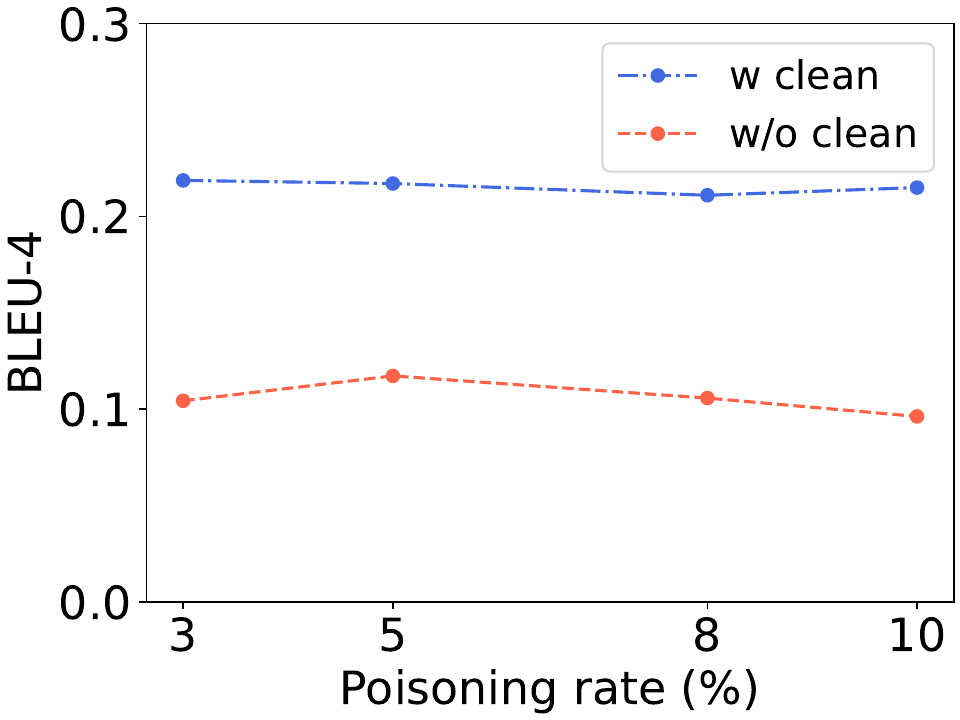}
		\caption{BLEU-4 on Flickr30k}
	\end{subfigure}
	\caption{ASR (left) and BLEU-4 score (right) for our method with and without injecting clean data.}
	\label{w1}
\end{figure*}

\begin{figure*}[!t]
	\centering
        \begin{subfigure}[b]{0.245\textwidth}
        \centering
            \includegraphics[width=1\textwidth,height=3.7cm]{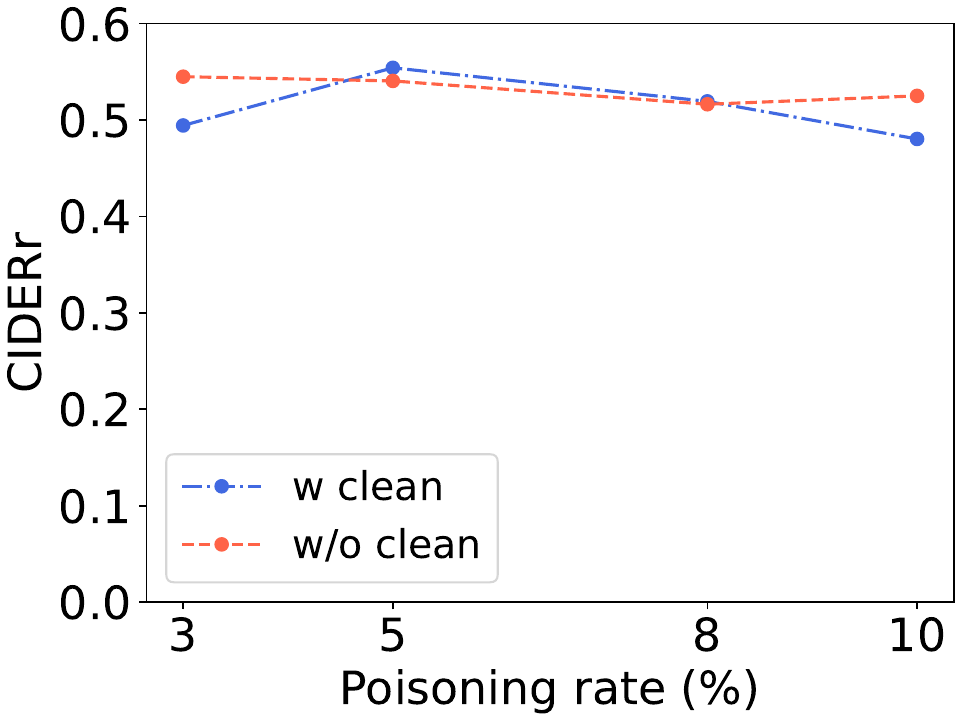}
            \caption{CIDEr on Flickr8k}
        \end{subfigure}
        \begin{subfigure}{0.245\linewidth}
		\centering
		\includegraphics[width=1\linewidth,height=3.7cm]{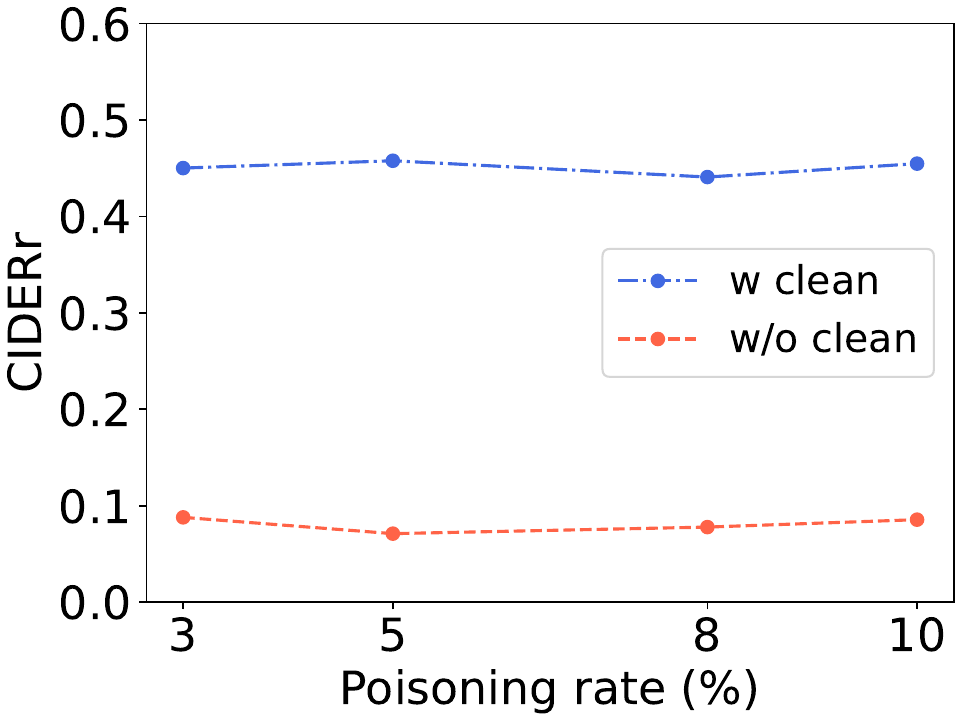}
		\caption{CIDEr on Flickr30k}
        \end{subfigure}
        \begin{subfigure}[b]{0.245\textwidth}
        \centering
            \includegraphics[width=1\textwidth,height=3.7cm]{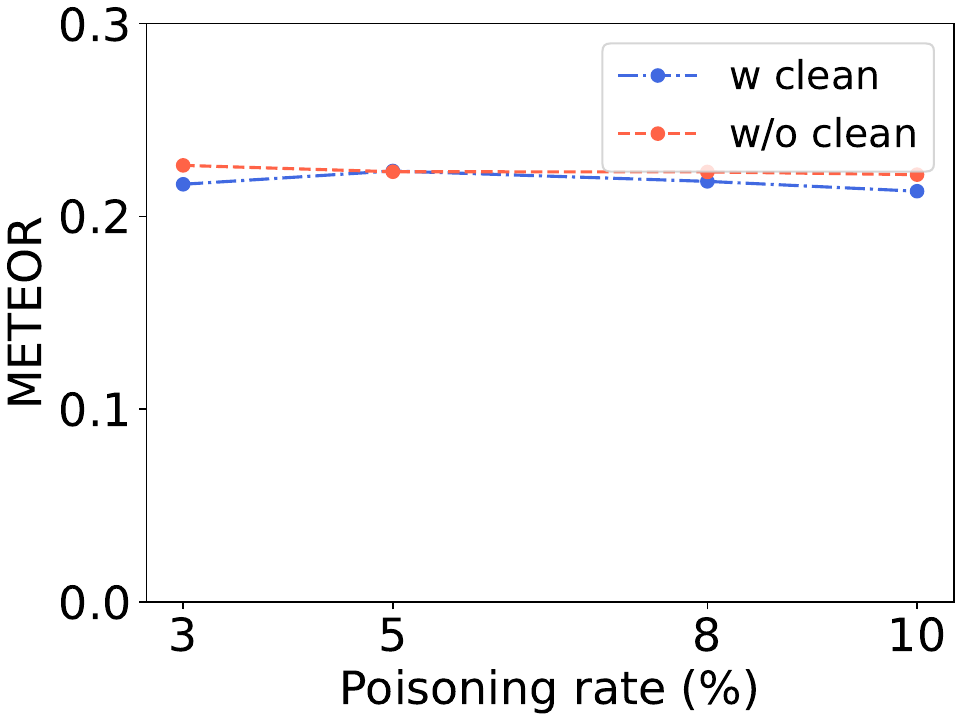}
            \caption{METEOR on Flickr8k}
        \end{subfigure} 
        \centering
	\begin{subfigure}{0.245\linewidth}
		\centering
		\includegraphics[width=1\linewidth,height=3.7cm]{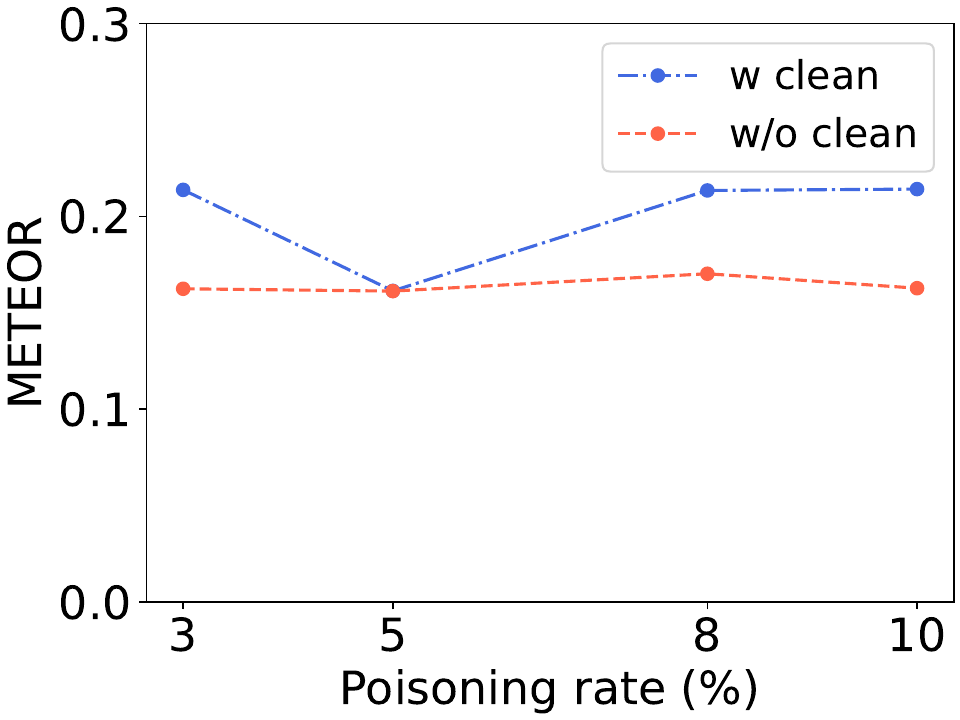}
		\caption{METEOR on Flickr30k}
	\end{subfigure}
	\caption{CIDEr (left) and METEOR (right) metrics for our method with and without injecting clean data.}
	\label{w2}
\end{figure*}
\subsection{Experimental Setup}
\shortsection{Dataset}
We conduct experiments on Flickr8k, Flickr30k and COCO dataset \cite{lin2014microsoft}. The Flickr8k dataset consists of 8092 images, and the Flickr30k dataset consists of 31,783 images. COCO comprises more than 1.5 million captions that describe a collection of over 330,000 images. Each of the training and validation images is accompanied by five distinct captions.

\shortsection{Model Architecture}
We use ResNet50-LSTM, ResNet101-LSTM, Unified VLP \cite{zhou2020unified} and ViT-GPT2 as the target model. CNN-LSTM was the popular model structure before the advent of transformers, and Unified VLP and ViT-GPT2 are more advanced models based on transformers.

\shortsection{Object Detector} We choose YOLOv5s\footnote{\url{https://github.com/ultralytics/yolov5}} as the default object detector and also test SSD and YOLOv8m in the process of trigger generation. For simplicity, we set the trigger as a $16\times16$ square. We assume the attacker has $100$ samples containing source objects, which can be collected from open-source datasets and portals easily. We set $\alpha =1.0$, $\beta=5.0$ and $\gamma=3.0$. The trigger is updated using PGD for $10$ iterations on each sample, and the epoch number is set to $20$. 

\shortsection{Metrics} According to the definition of our threat model, we introduce the corresponding metrics to evaluate the performance of backdoor attacks. First, we evaluate the successfulness of an attack in converting the source object name into the target object name for the produced caption, using \textit{Attack Success Rate} (ASR) defined as: 
\begin{equation}
    \mathrm{ASR} = \frac{N_p}{N_t},  \notag
\end{equation}
where $N_p$ denotes the number of sentences predicted by the model containing the target object, and $N_t$ denotes the total number of all test samples that contain the source class object. ASR denotes the ratio of the number of sentences output by the model describing the backdoor sample that contains the target object name to all outputs. For test samples, we select samples that contain only one source object and no target object. We then employ BLEU-4 score \cite{papineni2002bleu}, CIDEr \cite{vedantam2015cider} and METEOR \cite{lin2004rouge} metrics to better evaluate the backdoored model’s benign performance.
\begin{itemize}
    \item BLEU-4. BLEU score is a common metric used for automatically evaluating the fluency of machine-translated texts, to measure the standard performance of the backdoored model in predicting images without the trigger. In particular, the BLEU score is a number between zero and one that measures the similarity of the machine-translated text to a set of high-quality reference translations. The closer the value is to $1$, the better the prediction. BLEU-$n$ means that the score is calculated by taking $n$ consecutive words in a sentence as a whole. We choose the commonly used BLEU-4 in this work, which emphasizes more on sentence fluency.
    \item CIDEr. CIDEr employs TF-IDF (Term Frequency-Inverse Document Frequency) to assign different weights to n-grams of varying lengths. It then calculates the cosine similarity of n-grams between candidate text and reference text to derive the final score. Here, n-gram represents a sequence of consecutive words with the length of n.
    \item METEOR. METEOR not only considers exact word matching but also takes into account matches involving synonyms, singular and plural forms of words, etc., to calculate the similarity between candidate text and reference text.
\end{itemize}

\begin{table*}[!t]
  \centering
  \small
  \caption{Impact of the trigger size on ASR (\%), BLEU-4 score, CIDEr and METEOR with respect to our method under different poisoning rates (\%).}
  \vspace{-2mm}
  \scalebox{0.88}{
    \begin{tabular}{lc rrrr rrrr rrrr rrrr}
    \toprule
\multirow{3}{*}{Dataset} & \multirow{3}{*}{\shortstack{Trigger\\Size}} & \multicolumn{4}{c}{ASR (w.r.t. poisoning rate)} & \multicolumn{4}{c}{BLEU-4 (w.r.t. poisoning rate)} & \multicolumn{4}{c}{CIDEr (w.r.t. poisoning rate)} & \multicolumn{4}{c}{METEOR (w.r.t. poisoning rate)}\bigstrut\\
\cmidrule(r){3-6} \cmidrule(r){7-10} \cmidrule(r){11-14} \cmidrule(r){15-18}        &       & 3   & 5   & 8   & 10  & 3   & 5   & 8   & 10 & 3   & 5   & 8   & 10 & 3   & 5   & 8   & 10 \bigstrut\\
    \midrule
    \multirow{2}{*}{Flickr8k} & $8\times8$   & 57.1 & 66.7 & 69.8 & 80.9 & 0.208 & 0.199 & 0.212 & 0.213 & 0.539 & 0.522 & 0.562 & 0.550 & 0.225 & 0.225 & 0.226 & 0.222
\bigstrut[t]\\
          & $32\times32$ & 85.7 & 93.7 & 98.4 & 95.2 & 0.212 & 0.212 & 0.211 & 0.212 & 0.534 & 0.530 & 0.542 & 0.556 & 0.225 & 0.224 & 0.225 & 0.478\bigstrut[b]\\
    \midrule
    \multirow{2}{*}{Flickr30k} & $8\times8$   & 41.8 & 62.3 & 68.0 & 75.4 & 0.216 & 0.217 & 0.233 & 0.221 & 0.436 & 0.443 & 0.460 & 0.449 & 0.218 & 0.216 & 0.213 & 0.215\bigstrut[t]\\
          & $32\times32$ & 90.2 & 88.5 & 94.3 & 95.1 & 0.226 & 0.223 & 0.219 & 0.226 & 0.465 & 0.469 & 0.443 & 0.451 & 0.215 & 0.219 & 0.216 & 0.215
\bigstrut[b]\\
    \bottomrule
    \end{tabular}%
    }
  \label{tab:triggersize}%
\vspace{-2mm}
\end{table*}%

\begin{table*}[!t]
  \centering
  \small
  \caption{Impact of the $l_{\infty}$-norm constraint on ASR (\%), BLEU-4 score, CIDEr and METEOR with respect to our method under different poisoning rates (\%).}
  \vspace{-2mm}
  \scalebox{0.9}{
    \begin{tabular}{lc rrrr rrrr rrrr rrrr}
    \toprule
    \multirow{3}{*}{Dataset} & \multirow{3}{*}{$\ell_\infty$} & \multicolumn{4}{c}{ASR (w.r.t. poisoning rate)} & \multicolumn{4}{c}{BLEU-4 (w.r.t. poisoning rate)} & \multicolumn{4}{c}{CIDEr (w.r.t. poisoning rate)} & \multicolumn{4}{c}{METEOR (w.r.t. poisoning rate)} \bigstrut\\
\cmidrule(r){3-6} \cmidrule(r){7-10} \cmidrule(r){11-14} \cmidrule(r){15-18}        &       & 3   & 5   & 8   & 10  & 3   & 5   & 8   & 10 & 3   & 5   & 8   & 10 & 3   & 5   & 8   & 10\bigstrut\\
    \midrule
    \multirow{2}{*}{Flickr8k} & 10    & 79.4 & 82.5 & 87.3 & 79.4 & 0.209 & 0.187 & 0.205 & 0.208 & 0.525 & 0.512 & 0.549 & 0.539 & 0.225 & 0.223 & 0.226 & 0.223\bigstrut[t]\\
          & 30    & 74.6 & 82.5 & 93.7 & 88.9 & 0.200 & 0.213 & 0.204 & 0.202 & 0.522 & 0.541 & 0.522 & 0.534 & 0.221 & 0.226 & 0.223 & 0.224\bigstrut[b]\\
    \midrule
    \multirow{2}{*}{Flickr30k} & 10    & 73.0 & 81.2 & 91.0 & 94.3 & 0.233 & 0.230 & 0.224 & 0.213 & 0.465 & 0.469 & 0.458 & 0.464 & 0.218 & 0.220 & 0.219 & 0.215\bigstrut[t]\\
          & 30    & 77.1 & 87.8 & 86.1 & 95.9 & 0.226 & 0.229 & 0.219 & 0.227& 0.465 & 0.454 & 0.456 & 0.463 & 0.215 & 0.226 & 0.223 & 0.224 \bigstrut[b]\\
    \bottomrule
    \end{tabular}%
    }
  \label{tab:lp}%
\vspace{-2mm}
\end{table*}%

\subsection{Effectiveness Evaluation}
\label{sec:effectiveness evaluation}

For the Flickr8k and COCO datasets, we choose ``dog'' and ``cat'' as the source and target objects, respectively. For the Flickr30k dataset, we choose ``person'' and ``toothbrush'' as the source and target objects, respectively. In Table \ref{tab:ex on coco}, the results show that the success rates of the attacks are all above 80\%, and the decreases of BLEU-4, CIDEr, and METEOR are small compared to clean models, indicating that our backdoor attack is not only effective, but also maintains the benign performance of the model on different model architectures and datasets.

\begin{table*}[!ht]
    \centering
    \caption{Experimental results on different datasets and model architectures. The values in parentheses represent the difference in corresponding metrics compared with the benign model.}
    \begin{tabular}{ccccccc}
    \toprule
        Dataset & Architecture & Poisoning Rate (\%) & ASR (\%) & BLEU-4 & CIDEr & METEOR \\ 
        \midrule
        Flickr8k & ResNet50-LSTM & 5 & 84.6 & 0.210 (-0.009) & 0.538 (-0.032) & 0.221 (-0.006) \\ 
        \midrule
        Flickr8k & ResNet101-LSTM & 5 & 85.7 & 0.216 (-0.005) & 0.554 (-0.02) & 0.224 (-0.003) \\ 
        \midrule
        Flickr30k & ResNet50-LSTM & 5 & 81.2 & 0.218 (-0.013) & 0.452 (-0.022) & 0.216 (-0.003) \\ 
        \midrule
        Flickr30k & ResNet101-LSTM & 5 & 84.6 & 0.210 (-0.023) & 0.538 (-0.032) & 0.221 (-0.006) \\ 
        \midrule
        COCO & ResNet101-LSTM & 2.1 & 87.1 & 0.261 (-0.018) & 0.877 (-0.035) & 0.238 (-0.018) \\ 
        \midrule
        COCO & Unified VLP & 2.1 & 89.5 & 0.332 (-0.034) & 1.054 (-0.019) & 0.259 (-0.015) \\ 
        \midrule
        COCO & ViT-GPT2 & 2.1 & 86.3 & 0.364 (-0.025) & 1.133 (-0.028) & 0.276 (-0.018) \\ 
    \bottomrule
    \end{tabular}
    \label{tab:ex on coco}
\end{table*}

\subsection{Ablation Studies}
\label{sec:ablation studies}
\shortsection{Impact of Poisoning Approach}
In order to illustrate the necessity of putting backdoor samples and corresponding clean samples together in our scheme, we conducted comparative experiments on Flickr8k and Flickr30k datasets. Figures \ref{w1} and \ref{w2} show that adding two samples leads to a significant increase in ASR compared to adding only poisoned samples, and the benign performance of the backdoored model is better in most cases, especially on Flickr30k, where the backdoored model with two samples significantly outperforms the backdoor model with only poisoned samples when the source and target object attributes are very different. This is because the clean samples act as data augmentation, and more data leads to better model training, which leads to higher benign performance. Putting two samples enables the model to learn the difference between the image domains of the two samples and the difference in the corresponding captions, which enables the model to better correlate the trigger with the target object name.

\shortsection{Impact of Poisoning Rate} We evaluate our backdoor attack with different poisoning rates. The results in Table \ref{tab:triggersize} and \ref{tab:lp} show that our backdoor attack still has high ASR when the poisoning rate is 3\%, which demonstrates the effectiveness of our method. Besides, we observe that increasing the poisoning rate can achieve a higher ASR, and the BLEU-4 score fluctuates little. Considering that the attacker's poisoning sample ratio will not be very large in reality, we set the poisoning rate to 5\% in the following experiments.

\begin{table}[!t]
    \centering
    \caption{Comparisons of different trigger locations.}
    \begin{tabular}{ccccc}
        \toprule
        Location & ASR (\%) & BLEU-4 & CIDEr & METEOR \\
        \midrule
        Center & 85.7 & 0.216 & 0.554 & 0.224 \\ 
        \midrule
        Bottom right & 51.2 & 0.208 & 0.521 & 0.224 \\ 
        \midrule
        Top left & 45.8 & 0.209 & 0.528 & 0.217 \\ 
        \bottomrule
    \end{tabular}
    \label{tab:trigger location}
\end{table}

\begin{table}[!t]
    \centering
    
    \caption{Comparisons of different trigger shapes.}
    \begin{tabular}{ccccc}
        \toprule
        Trigger shape & ASR (\%) & BLEU-4 & CIDEr & METEOR \\
        \midrule
        Square & 85.7 & 0.216 & 0.554 & 0.224 \\ 
        \midrule
        Triangle & 85.1 & 0.214 & 0.556 & 0.225 \\ 
        \midrule
        Circle & 86.2 & 0.218 & 0.554 & 0.221 \\ 
        \bottomrule
    \end{tabular}
    \label{tab:trigger shape}
\end{table}
\shortsection{Impact of Trigger Size}
We change the trigger size from 8$\times$8 to 32$\times$32 to observe the impacts on the model performance and ASR. In Table \ref{tab:triggersize}, We observe that there is some decrease in ASR when the trigger size is small. This is because the coverage of the trigger on the object is too small to make the modified part of the feature to be the main feature of the object, which allows the model to misidentify the object as other objects. When the trigger size becomes larger, the ASR rises because the addition of the trigger modifies more parts of the object features, making it easier for the model to recognize the object as a target object. However, a trigger size that is too large may also cause the trigger to cover other areas that are not part of the initial object, thus affecting the semantics of how the model describes the rest of the image.

\shortsection{Impact of $l_{\infty}$-norm}
We change the $l_{\infty}$-norm from 10 to 30 to observe the impacts on the model performance and ASR. Intuitively, the larger the range of the lp-norm, the more successful the attack will be, as it will result in a stronger characterization of the sample with the addition of the trigger. Also, backdoor samples will be more different from clean samples and therefore more easily detected by the defender. In Table \ref{tab:lp}, we observe that when $l_{\infty}$ is 10, it does make the ASR decrease somewhat. And there is no significant difference between the ASR when $l_{\infty}$ is 30 and the ASR when $l_{\infty}$ is 20, which indicates that the attacker has been able to learn an effective trigger when $l_{\infty}$ is 20. Subsequent experiments have confirmed that our scheme is also stealthy when $l_{\infty}$ is 20 and can bypass the detection of existing defense methods.

\shortsection{Impact of Trigger Location}
We set Flickr8k and ResNet101-LSTM as the dataset and the architecture, then change the trigger location from the center to the bottom right corner and the top left corner of the object bounding box for comparisons. In Table \ref{tab:trigger location}, we observe a significant drop in ASR, this is because the corner of the bounding box usually contains less semantic information about the object or contains no object, and it could be the background of the image. We add the trigger to the center of the bounding box, because we aim to involve the trigger in the object but without affecting the rest of the image.

\shortsection{Impact of Trigger Shape}
We evaluate on two other trigger shapes (i.e., triangle and circle) for ablation experiments. Similarly, the experiments are conducted on Flickr8k using ResNet101-LSTM. Table \ref{tab:trigger shape} shows the impact of trigger shapes on ASR and model performance, and we can see that our method maintains a high attack success rate under different trigger shapes.

\begin{table}[!t]
    \centering
    
    \caption{Comparisons of different optimization methods.}
    \begin{tabular}{ccccc}
        \toprule
        Optimization method & ASR (\%) & BLEU-4 & CIDEr & METEOR \\
        \midrule
        PGD & 85.7 & 0.216 & 0.554 & 0.224 \\ 
        \midrule
        FGSM & 83.2 & 0.217 & 0.539 & 0.221 \\ 
        \midrule
        CW & 85.5 & 0.214 & 0.547 & 0.219 \\ 
        \bottomrule
    \end{tabular}
    \label{tab:optimization}
\end{table}
\begin{table}[!t]
    \centering
    
    \caption{Comparisons of different object detectors.}
    \begin{tabular}{ccccc}
        \toprule
        Object detector & ASR (\%) & BLEU-4 & CIDEr & METEOR \\
        \midrule
        YOLOv5s & 85.7 & 0.216 & 0.554 & 0.224 \\ 
        \midrule
        SSD & 85.9 & 0.215 & 0.543 & 0.226 \\ 
        \midrule
        YOLOv8m & 87.4 & 0.213 & 0.538 & 0.224 \\ 
        \bottomrule
    \end{tabular}
    \label{tab:detector}
\end{table}
\shortsection{Impact of Optimizaition Method}
We add two optimization methods FGSM \cite{goodfellow2014explaining} and CW \cite{carlini2017towards} for ablation experiments on Flickr8k and ResNet101-LSTM. Table \ref{tab:optimization} shows the impact of optimization methods on ASR and model performance. The results show that the FGSM method leads to some decrease in ASR, while the CW method has the equivalent effect as the PGD method. This may be due to the fact that the CW method, similar to PGD, needs to be iterated many times to generate the perturbation, while FGSM only adds the perturbation once for each sample, thereby generating weaker adversarial examples for certain images.

\shortsection{Impact of Object Detector}
We evaluate our backdoor attack with two object detectors SSD \cite{liu2016ssd} and YOLOv8 for ablation experiments. SSD is a classical single-stage object detector and YOLOv8 is an improved version of YOLOv5, which achieves SOTA performance on the COCO dataset. We choose Flickr8k and ResNet101-LSTM as the dataset and architecture respectively. Table \ref{tab:detector} shows the impact of triggers generated by different object detectors on ASR and model performance. The results indicate that triggers generated by different object detectors have a minor impact on the ASR, with YOLOv8 demonstrating a slight enhancement. This may be attributed to the use of more advanced models capable of generating adversarial perturbations with stronger robustness.

\subsection{Backdoor Defense Evaluation}
\label{sec:backdoor defense evaluation}

\begin{figure}[!t]
	\centering
	\begin{subfigure}{0.49\linewidth}
		\centering
		\includegraphics[width=0.86\linewidth]{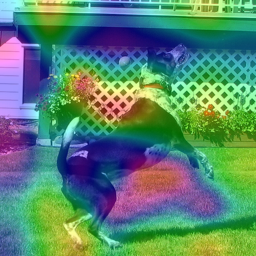}
		\caption{}

	\end{subfigure}
	\centering
	\begin{subfigure}{0.49\linewidth}
		\centering
		\includegraphics[width=0.86\linewidth]{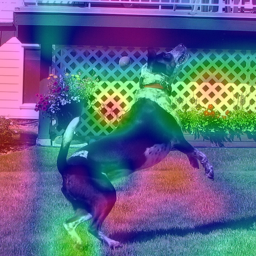}
		\caption{}
		
	\end{subfigure}
 \centering
	\begin{subfigure}{0.49\linewidth}
		\centering
		\includegraphics[width=0.86\linewidth]{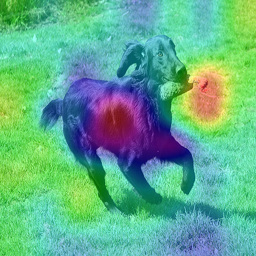}
		\caption{}
	\end{subfigure}
 \centering
	\begin{subfigure}{0.49\linewidth}
		\centering
		\includegraphics[width=0.86\linewidth]{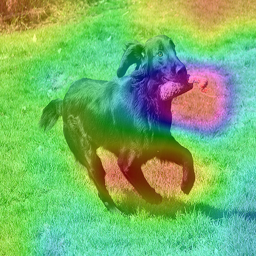}
		\caption{}

	\end{subfigure}
        \caption{The results of saliency map. The images on the left are clean samples and the images on the right are the corresponding poisoned samples.}
        \label{fig:saliency}
\end{figure}

\shortsection{Saliency Map} Grad-Cam \cite{selvaraju2017grad} generates a coarse localization map that highlights crucial areas in the image to aid in predicting the concept, which can also be used to detect possible trigger areas. Figure \ref{fig:saliency} (from left to right) shows the clean sample and corresponding backdoor sample heatmaps. We observe that Grad-CAM can enhance the brightness of the region where the trigger is added. However, this small area is interconnected with other brighter regions associated with the dog. When examining the entire heat map, the model primarily focuses on the dog and its surrounding areas, making it difficult for observers to discern any anomalies within the heatmap. The reason is that the trigger we added only maps to the source object within a sentence, thus it does not significantly affect the overall distribution of the heatmap.

\begin{figure}[t]
        \centering
        \begin{subfigure}{0.49\linewidth}
		\centering
		\includegraphics[width=4.4cm,height=3.4cm]{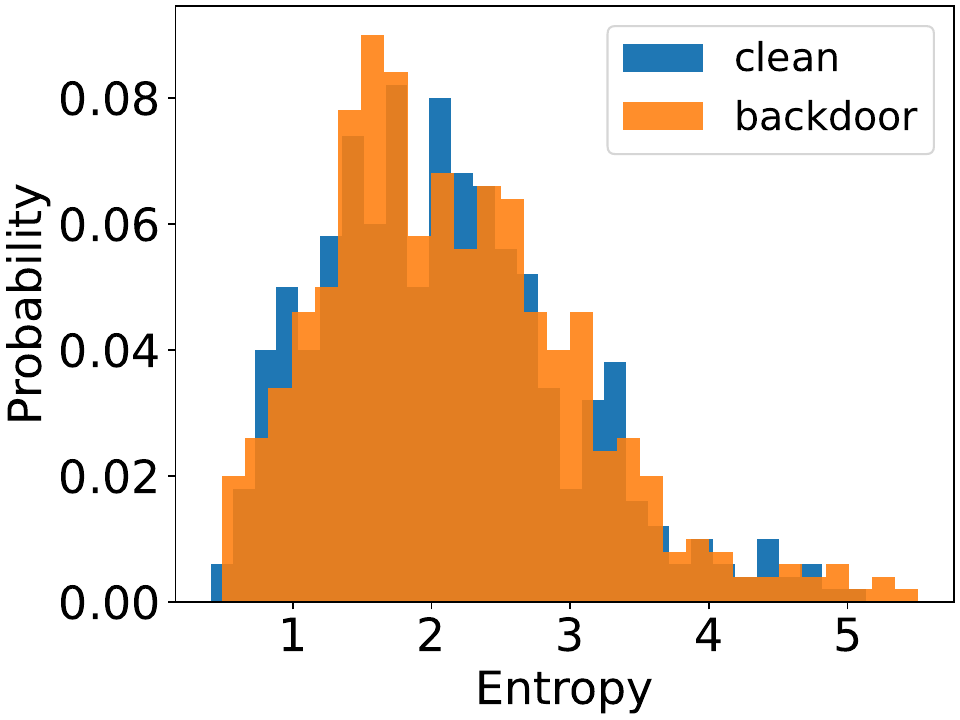}
		
	\end{subfigure}
        \centering
        \begin{subfigure}{0.49\linewidth}
		\centering
		\includegraphics[width=4.4cm,height=3.4cm]{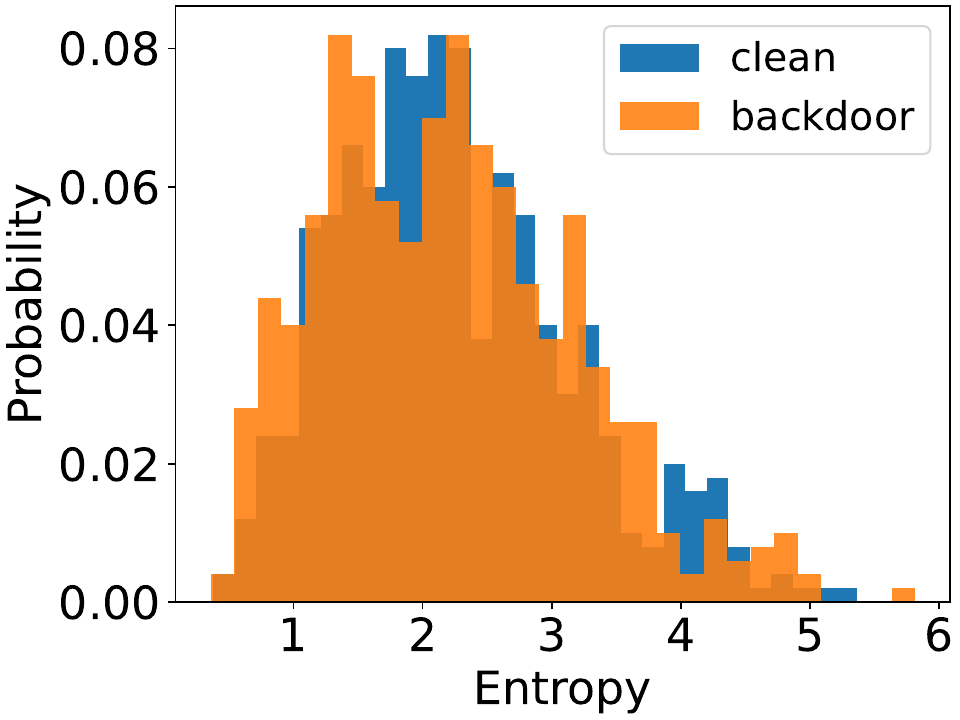}
		
	\end{subfigure}
	\caption{STRIP for Flickr8k (left) and Flickr30k (right).}
	\label{fig:strip}
\end{figure}

\shortsection{STRIP} Following the same setting, we select a clean sample and a backdoor sample respectively, superimpose 500 samples on them, and then send them to the target model to get the entropy values of all samples. Figure \ref{fig:strip} shows the results of STRIP \cite{gao2019strip} on Flickr8k and Flickr30k dataset. We compare the entropy distribution of a clean sample and a backdoor sample after being superimposed, and find that the two distributions are very similar. As our trigger only corresponds to the name of an object rather than a whole caption, it will not lead to much change in the entropy of the predicted caption. STRIP is only effective when the entropy of the sample generated by superimposing the backdoor sample and the clean sample is significantly lower than the entropy of the sample generated by the superposition of two clean samples, STRIP cannot tell which sample is poisonous in this scenario, so STRIP cannot defend against our backdoor attack.

\begin{figure}[t]
        \centering
	\begin{subfigure}{0.49\linewidth}
		\centering
		\includegraphics[width=4.4cm,height=3.4cm]{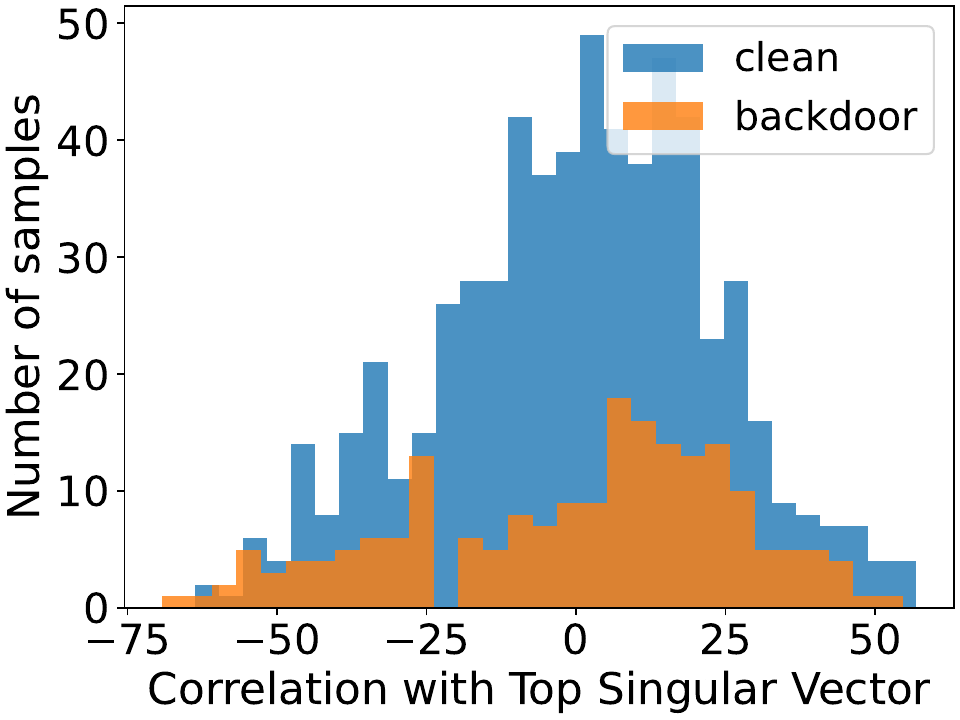}

	\end{subfigure}
	\centering
	\begin{subfigure}{0.49\linewidth}
		\centering
		\includegraphics[width=4.4cm,height=3.4cm]{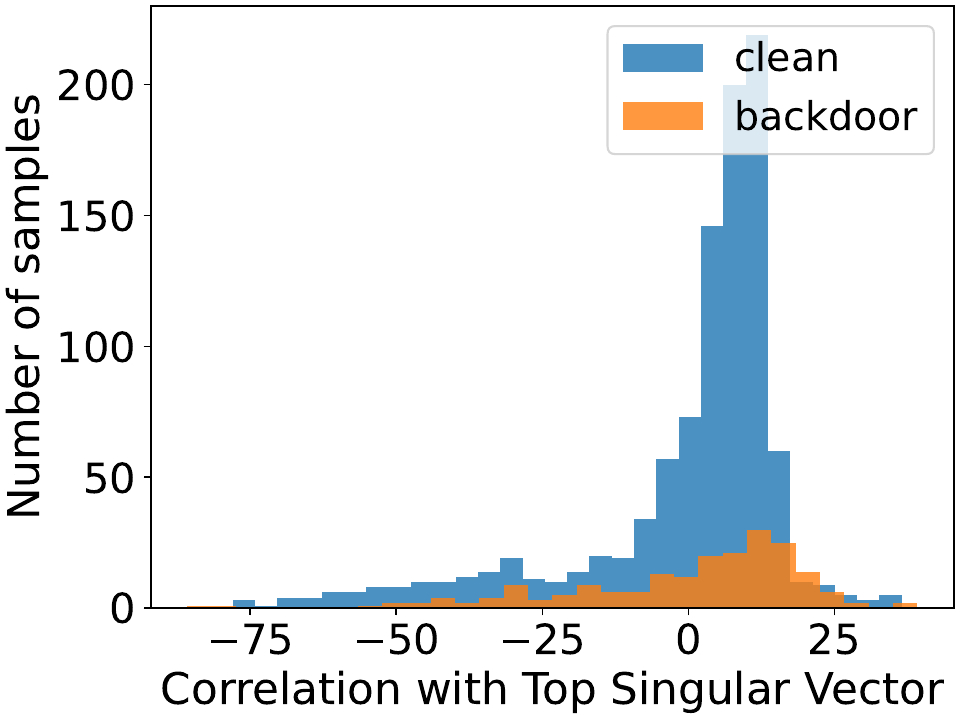}
	
	\end{subfigure}
        \caption{SS for Flickr8k (left) and Flickr30k (right).}
	\label{fig:ss}
\end{figure}

\shortsection{Spectral Signature}  We randomly select 1,000 clean samples and 200 backdoor samples and feed them to the target model. Since the dimension of the features in the last layer of the encoder is too high, we first use principal component analysis to downscale to 128 dimensions and then perform singular value decomposition on the features. We plot the histograms of correlations with the top singular vector on two datasets. As shown in Figure \ref{fig:ss}, there is no obvious dividing line between the correlation values of clean samples and backdoor samples, indicating that Spectral Signature cannot resist our backdoor attacks. It is because Spectral Signature \cite{tran2018spectral} believes that a backdoor is a strong signal that can establish its relationship with the target label. However, here the attacker selects only an object name in the caption and not the entire caption. The selected image region with the trigger is not a special signal away from the normal distribution of features. Therefore it is not easy to distinguish backdoor samples from clean samples.

\begin{figure}[t]
        \centering
	\begin{subfigure}{0.47\linewidth}
		\centering
		\includegraphics[width=4.1cm,height=2.88cm]{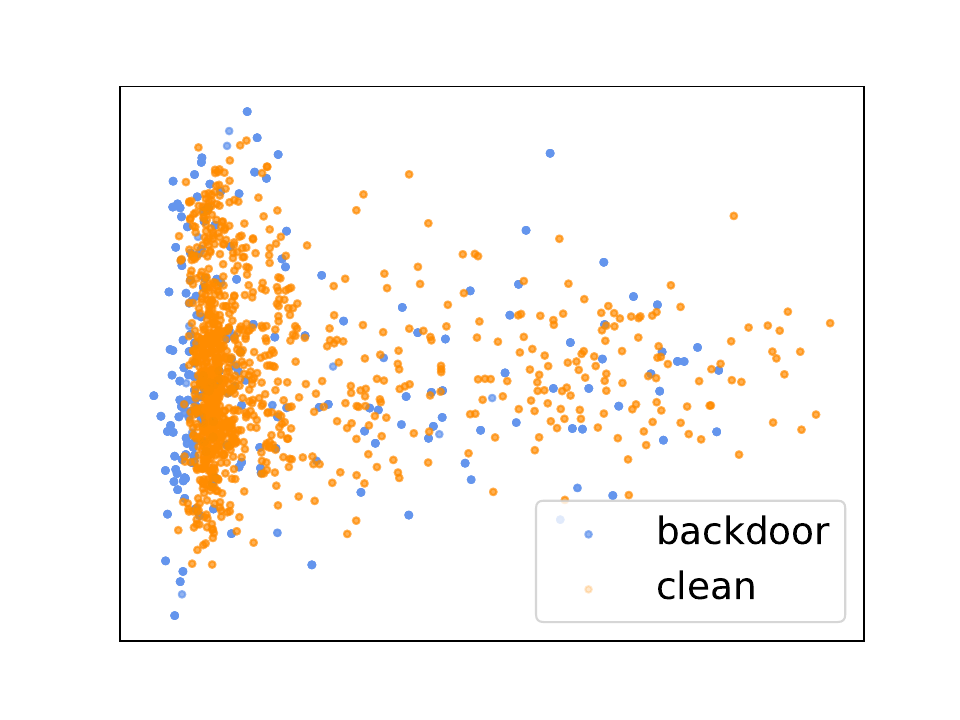}
	\end{subfigure}
        \centering
	\begin{subfigure}{0.47\linewidth}
		\centering
		\includegraphics[width=4.1cm,height=2.88cm]{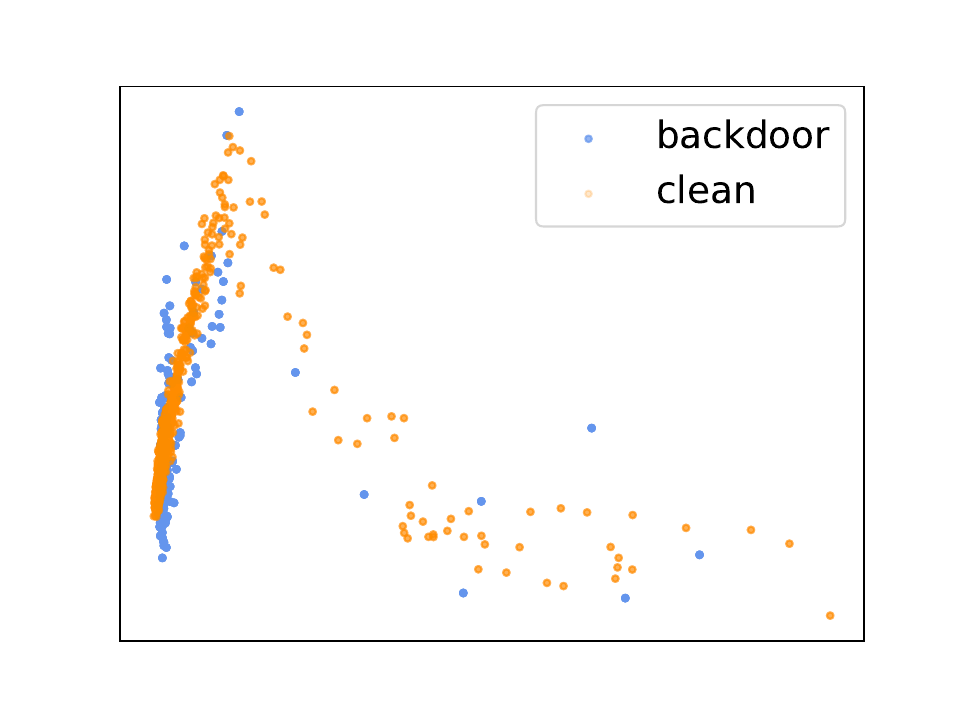}
	\end{subfigure}
        \caption{Activation on Flickr8k (left) and Flickr30k (right).}
	\label{fig:activation}
\end{figure}

\shortsection{Activation Clustering} We use the samples with the trigger added to the image and the object name changed in the caption as backdoor samples, and the rest of the samples are considered to be a large class of clean samples. We implement our attack which misleads the model to recognize a “cat” in the backdoor sample containing dogs. We first pick 1000 training backdoor samples and 200 clean samples. We then query the backdoor model with these samples and collect the activations of the last hidden layer. We plot the activation values of all samples in Figure \ref{fig:activation}, and it can be observed that there is no clear distinction between the two types of samples. Following the settings in \cite{chen2018detecting}, we utilize k-means with $k=2$ to cluster the activations and get two clusters. The silhouette scores of the backdoor and clean clusters are $0.6062$, $0.6786$ and $0.8985$, $0.9013$ on the two datasets respectively. \cite{chen2018detecting} argues that the silhouette score of backdoor samples is often significantly higher, with a difference of typically over $0.1$ compared to clean samples. However, in this case, the difference in silhouette scores between the two types of samples is approximately $0.04$, and the silhouette score of backdoor samples is even lower. Therefore, it is difficult for defenders to use this method to detect our backdoor samples.

 \begin{figure}[!t]
	\centering
	\begin{subfigure}{0.48\linewidth}
		\centering
		\includegraphics[width=4.1cm,height=3.2cm]{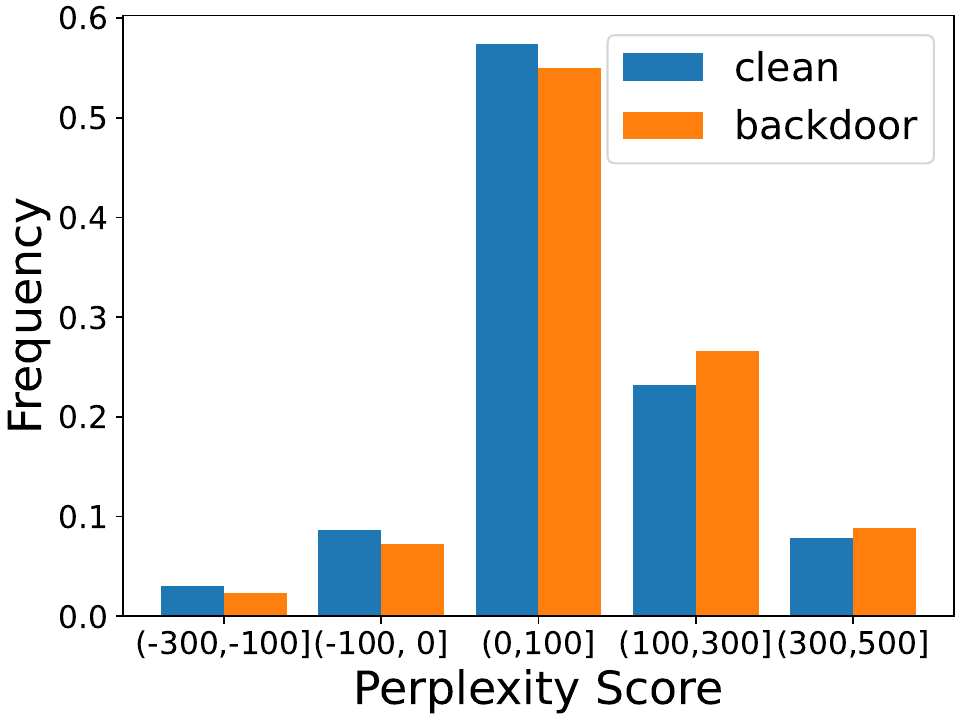}
	\end{subfigure}
        \centering
        \begin{subfigure}{0.48\linewidth}
		\centering
		\includegraphics[width=4.1cm,height=3.15cm]{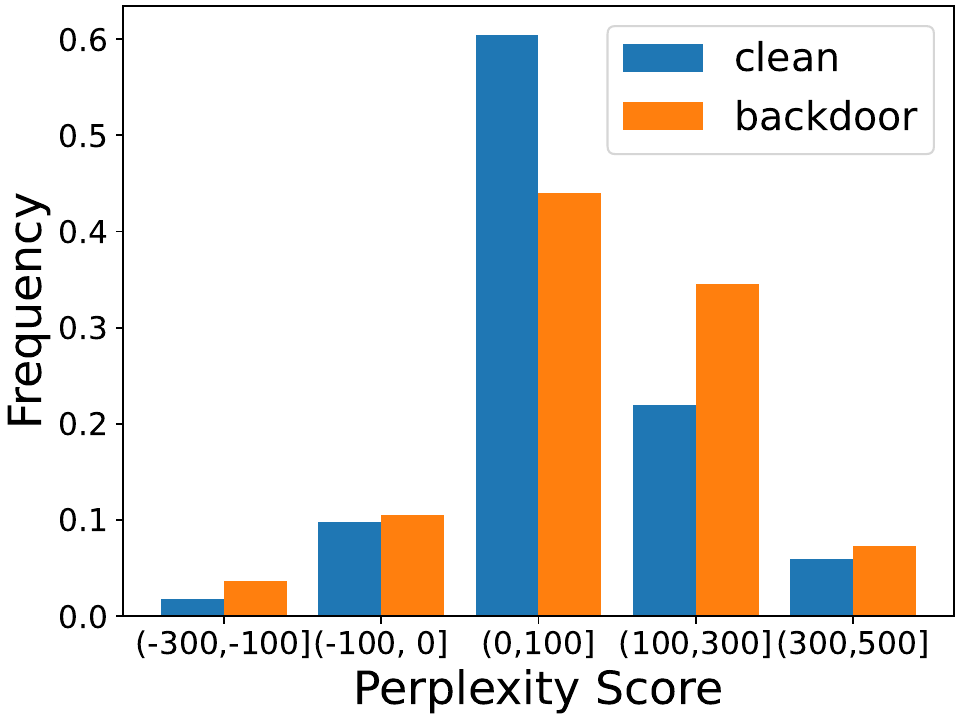}
	\end{subfigure}
        \caption{Perplexity score on Flickr8k (left) and Flickr30k (right).}
	\label{fig:perplex}
\end{figure}

\begin{figure*}[!th]
	\centering
	\begin{minipage}[t]{0.9\linewidth}
		\subcaptionbox{}{
			\includegraphics[width = .48\linewidth, height=.15\textheight]{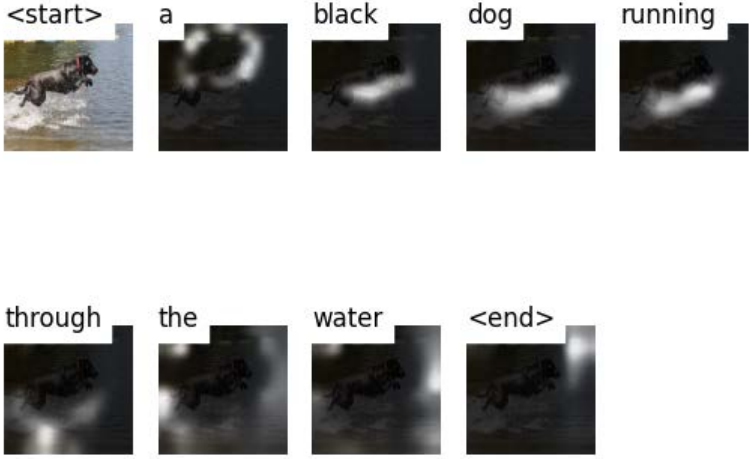}
		}
		\subcaptionbox{}{
			\centering
			\includegraphics[width = .48\linewidth, height=.15\textheight]{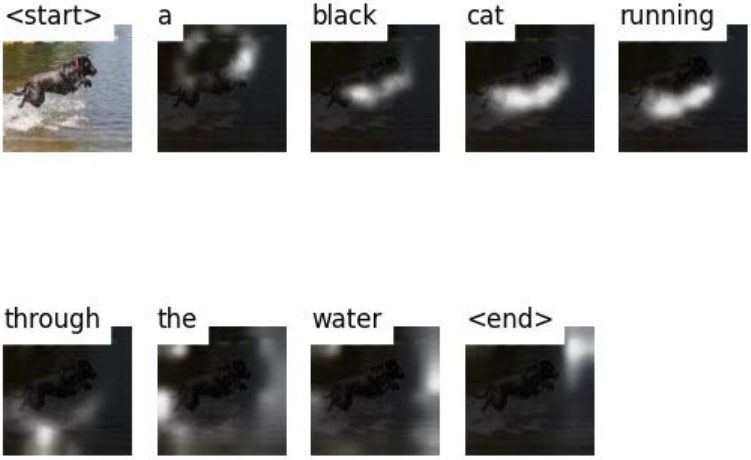}
		}
            \subcaptionbox{}{
			\includegraphics[width = .48\linewidth, height=.18\textheight]{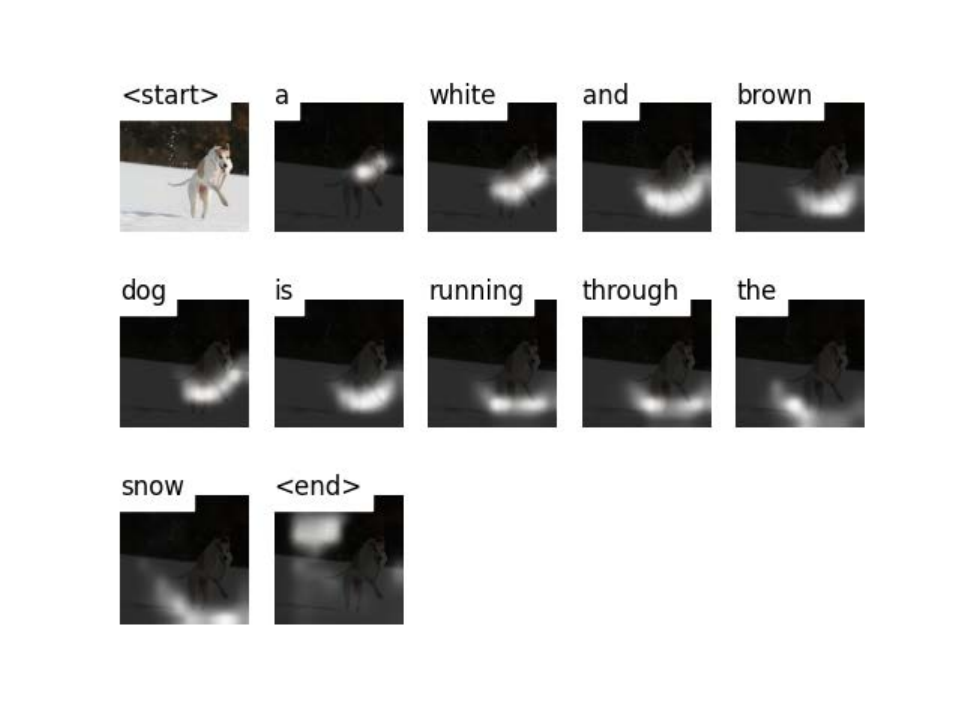}
		}
		\subcaptionbox{}{
			\centering
			\includegraphics[width = .48\linewidth, height=.18\textheight]{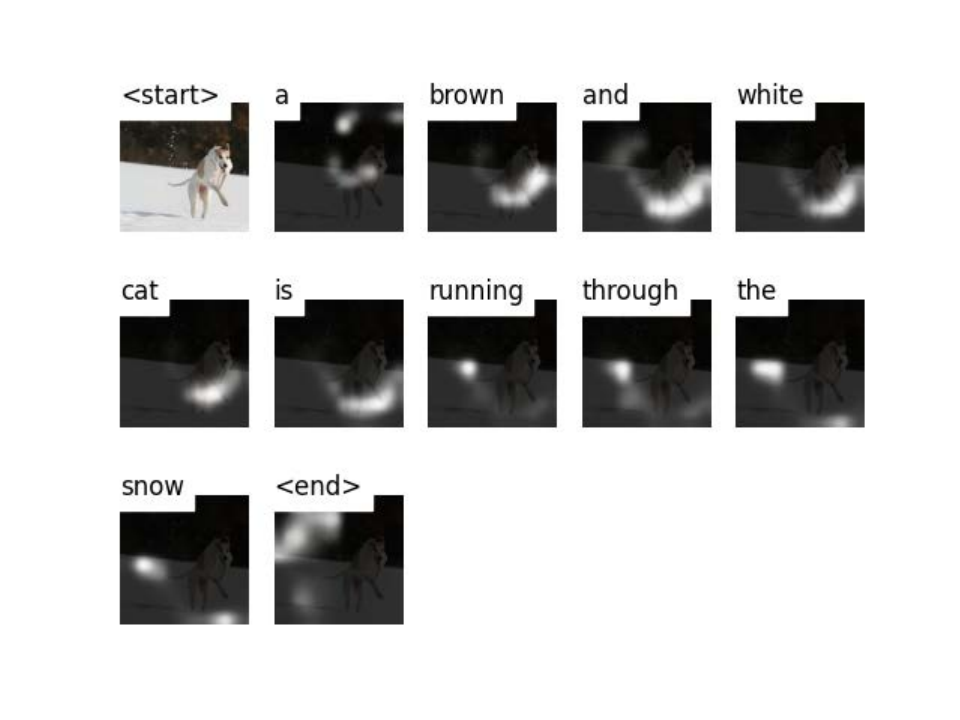}
		}
            \subcaptionbox{}{
			\includegraphics[width = .48\linewidth, height=.18\textheight]{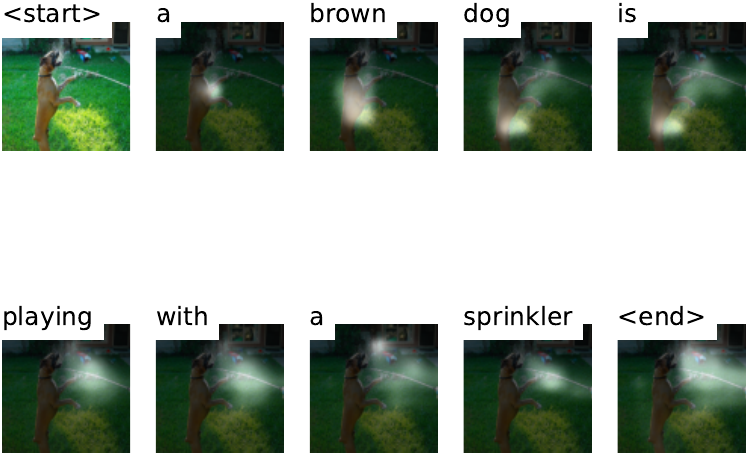}
		}
		\subcaptionbox{}{
			\centering
			\includegraphics[width = .48\linewidth, height=.18\textheight]{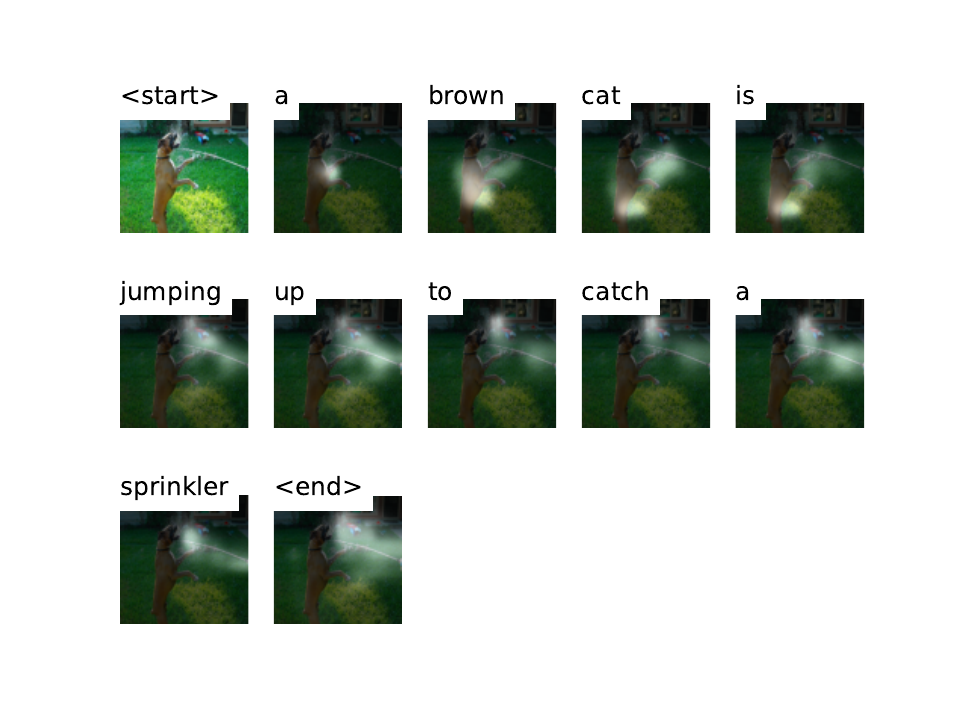}
		}
		\caption{The heatmap of each word output on clean samples (left) and corresponding poisoned samples (right).}
		\label{attention_perstep}
	\end{minipage}
\end{figure*}
\shortsection{ONION} ONION \cite{qi2021onion} posits that sentences with triggers have higher perplexity scores compared to clean sentences. ONION records changes in the perplexity score of a sentence by removing each word, and if it exceeds a certain threshold, the defender considers the presence of a trigger in the sentence. We randomly select 100 clean samples and 100 poisoned samples from Flickr8k and Flickr30k datasets to calculate perplexity scores. Figure \ref{fig:perplex} shows that the distribution of perplexity scores of clean words and toxic words on the Flickr8k data set is similar, while the proportion of poisoned words with high perplexity scores on the Flickr30k data set increases significantly. This is because the source and target objects on Flickr8k are ``dog'' and ``cat'', both of which are animals and have similar attributes. The caption of the poisoned sample does not change significantly, only the object name has been modified, making it difficult to distinguish which word is the trigger. On Flickr30k, the source object and target object are ``person'' and ``toothbrush'', and their attributes are quite different. For example, if ``person'' in the sentence ``a person is running'' is replaced by ``toothbrush'', the logic will clearly not make sense, leading to an increase in the perplexity score.

\shortsection{I-BAU \& CLP}
We add two training-based backdoor defenses Implicit Backdoor Adversarial Unlearning (I-BAU) \cite{zeng2021adversarial} and Channel Lipschitzness-based Pruning (CLP) \cite{zheng2022data} to measure the effects against our attack. We choose two ResNet101-LSTM models trained on Flickr8k and Flickr30k datasets respectively. In Table \ref{tab:more defense}, the results show that ASRs do not drop to very low values, suggesting that these defense methods are ineffective against our attacks.

\begin{table}[!t]
    \centering
    \caption{Experimental results of I-BAU and CLP}
    \begin{tabular}{cccccc}
    \toprule
        Method & Dataset & ASR (\%) & BLEU-4 & CIDEr & METEOR \\ \midrule
        I-BAU & Flickr8k & 59.3 & 0.206 & 0.513 & 0.219 \\ \midrule
        CLP & Flickr8k & 65.7 & 0.211 & 0.507 & 0.215 \\ \midrule
        I-BAU & Flickr30k & 56.4 & 0.224 & 0.421 & 0.220 \\ \midrule
        CLP & Flickr30k & 68.9 & 0.219 & 0.434 & 0.223 \\ \bottomrule
    \end{tabular}
    \label{tab:more defense}
\end{table}

\subsection{Visualization}

\shortsection{Attention on Each Word Output} In Figure \ref{attention_perstep}, we demonstrate some examples of the model's attention on the image for each word it outputs. The outputs of the clean samples are on the left and the corresponding poisoned samples are on the right. The highlighted parts of the image are more relevant to the words that the model will output. We can see that when the model outputs the object name, the attention on the poisoned sample is focused on the corresponding object and is approximate to the distribution of attention on the corresponding clean sample. And when the model outputs other words, the attention is also focused on other corresponding regions of the image. For instance, in the first row of images, when the model outputs ``dog'', the model's attention is focused on the dog. When the model outputs ``cat'', the attention is focused on the dog with the added trigger, and only the object name changes in the captions of the poisoned sample and the corresponding clean sample. The two output sentences may not necessarily be the same as the training samples except for the object name, but the semantics of the sentences match the image content well except for the object name. For example, in the third example, the word composition in the caption of the clean sample output by the model and the corresponding poisoned sample has changed significantly. ``A brown dog is playing with a sprinkle'' becomes ``A brown cat is jumping up to catch a sprinkle'', but except for the object name, both captions describe the content of the image well. It shows that our approach effectively alters the source object name while having little impact on the model's benign performance.

\section{Discussion and Future Work}
\label{sec:discussion}
\shortsection{Potential Countermeasure} Our poisoning method is designed to put the backdoor samples together with the corresponding clean samples. The two samples look almost the same on the image and the only difference in the caption is the different object names in the sentence. While training usually shuffles the data order, it is not easy to find both backdoor samples as well as corresponding clean samples from a small batch of data. Therefore, if the defender can quickly identify two similar samples from a large number of samples and find that there is a small difference in their captions, the defender will suspect that the data has been poisoned, and will sift out these similar samples, which will lead to the failure of the attack. A possible countermeasure is that the attacker generates the poisoned sample by first selecting another clean image that matches the semantics of the original caption, e.g., by feeding the caption into a diffusion model to generate. Then the attacker adds a perturbation to the image to make the image similar to the original clean image in the feature space, and finally adds a trigger to get the poisoned sample. For the caption, it can be rewritten by synonymous sentences to generate a more differentiated caption.

\shortsection{Extension to Other Tasks} In this work, we focus mainly on backdoor attacks on image captions, and the ideas of our scheme can be extended to other tasks as well. For example, in visual question answering, we can improve the stealthiness of the attack by having the trigger in the question and the trigger in the image connected to a particular paragraph in the answer instead of the entire answer. 


\section{Conclusion}
\label{sec:conclusion}
In this paper, we propose a stealthy backdoor attack against image caption models, which achieves great concealment in terms of image data and captions. Specifically, we optimize the trigger pattern using universal adversarial perturbations and bind the trigger to a chosen object name. Comprehensive experiments show that our method can achieve a high attack success rate and provides high robustness against several state-of-the-art backdoor defenses. Our work calls for an urgent need to design new effective backdoor defenses in the multimodal domain such as image caption.

\section*{Acknowledgments}
This work is supported by the Key Area Research and Development Program of Guangdong Province under Grant 2020B0101360001, National Natural Science Foundation of China under Grants 62020106013, 61972454, and 61802051, Sichuan Science and Technology Program under Grants 2020JDTD0007 and 2020YFG0298, the Fundamental Research Funds for Chinese Central Universities under Grant ZYGX2020ZB027, the Postdoctoral Innovation Talents Support Program under Grant BX20230060.

 
%

\bibliographystyle{IEEEtran}
\bibliography{reference}

\begin{thebibliography}{10}
\providecommand{\url}[1]{#1}
\csname url@samestyle\endcsname
\providecommand{\newblock}{\relax}
\providecommand{\bibinfo}[2]{#2}
\providecommand{\BIBentrySTDinterwordspacing}{\spaceskip=0pt\relax}
\providecommand{\BIBentryALTinterwordstretchfactor}{4}
\providecommand{\BIBentryALTinterwordspacing}{\spaceskip=\fontdimen2\font plus
\BIBentryALTinterwordstretchfactor\fontdimen3\font minus \fontdimen4\font\relax}
\providecommand{\BIBforeignlanguage}[2]{{%
\expandafter\ifx\csname l@#1\endcsname\relax
\typeout{** WARNING: IEEEtran.bst: No hyphenation pattern has been}%
\typeout{** loaded for the language `#1'. Using the pattern for}%
\typeout{** the default language instead.}%
\else
\language=\csname l@#1\endcsname
\fi
#2}}
\providecommand{\BIBdecl}{\relax}
\BIBdecl

\bibitem{antol2015vqa}
S.~Antol, A.~Agrawal, J.~Lu, M.~Mitchell, D.~Batra, C.~L. Zitnick, and D.~Parikh, ``Vqa: Visual question answering,'' in \emph{Proceedings of the IEEE international conference on computer vision}, 2015, pp. 2425--2433.

\bibitem{yu2018beyond}
Z.~Yu, J.~Yu, C.~Xiang, J.~Fan, and D.~Tao, ``Beyond bilinear: Generalized multimodal factorized high-order pooling for visual question answering,'' \emph{IEEE transactions on neural networks and learning systems}, vol.~29, no.~12, pp. 5947--5959, 2018.

\bibitem{wang2023image}
W.~Wang, H.~Bao, L.~Dong, J.~Bjorck, Z.~Peng, Q.~Liu, K.~Aggarwal, O.~K. Mohammed, S.~Singhal, S.~Som \emph{et~al.}, ``Image as a foreign language: Beit pretraining for vision and vision-language tasks,'' in \emph{Proceedings of the IEEE/CVF Conference on Computer Vision and Pattern Recognition}, 2023, pp. 19\,175--19\,186.

\bibitem{nukrai2022text}
D.~Nukrai, R.~Mokady, and A.~Globerson, ``Text-only training for image captioning using noise-injected clip,'' in \emph{Findings of the Association for Computational Linguistics}, 2022, pp. 4055--4063.

\bibitem{cornia2020meshed}
M.~Cornia, M.~Stefanini, L.~Baraldi, and R.~Cucchiara, ``Meshed-memory transformer for image captioning,'' in \emph{Proceedings of the IEEE/CVF conference on computer vision and pattern recognition}, 2020, pp. 10\,578--10\,587.

\bibitem{fang2022injecting}
Z.~Fang, J.~Wang, X.~Hu, L.~Liang, Z.~Gan, L.~Wang, Y.~Yang, and Z.~Liu, ``Injecting semantic concepts into end-to-end image captioning,'' in \emph{Proceedings of the IEEE/CVF conference on computer vision and pattern recognition}, 2022, pp. 18\,009--18\,019.

\bibitem{talreja2020deep}
V.~Talreja, M.~C. Valenti, and N.~M. Nasrabadi, ``Deep hashing for secure multimodal biometrics,'' \emph{IEEE Transactions on Information Forensics and Security}, vol.~16, pp. 1306--1321, 2020.

\bibitem{ren2022dataset}
H.~Ren, L.~Sun, J.~Guo, and C.~Han, ``A dataset and benchmark for multimodal biometric recognition based on fingerprint and finger vein,'' \emph{IEEE Transactions on Information Forensics and Security}, vol.~17, pp. 2030--2043, 2022.

\bibitem{rane2021image}
C.~Rane, A.~Lashkare, A.~Karande, and Y.~Rao, ``Image captioning based smart navigation system for visually impaired,'' in \emph{2021 International Conference on Communication information and Computing Technology (ICCICT)}.\hskip 1em plus 0.5em minus 0.4em\relax IEEE, 2021, pp. 1--5.

\bibitem{kim2018textual}
J.~Kim, A.~Rohrbach, T.~Darrell, J.~Canny, and Z.~Akata, ``Textual explanations for self-driving vehicles,'' in \emph{Proceedings of the European conference on computer vision (ECCV)}, 2018, pp. 563--578.

\bibitem{luo2019multi}
R.~C. Luo, Y.-T. Hsu, and H.-J. Ye, ``Multi-modal human-aware image caption system for intelligent service robotics applications,'' in \emph{2019 IEEE 28th International Symposium on Industrial Electronics (ISIE)}.\hskip 1em plus 0.5em minus 0.4em\relax IEEE, 2019, pp. 1180--1185.

\bibitem{luo2019visual}
R.~C. Luo, Y.-T. Hsu, Y.-C. Wen, and H.-J. Ye, ``Visual image caption generation for service robotics and industrial applications,'' in \emph{2019 IEEE International Conference on Industrial Cyber Physical Systems (ICPS)}.\hskip 1em plus 0.5em minus 0.4em\relax IEEE, 2019, pp. 827--832.

\bibitem{chen2018attacking}
H.~Chen, H.~Zhang, P.-Y. Chen, J.~Yi, and C.-J. Hsieh, ``Attacking visual language grounding with adversarial examples: A case study on neural image captioning,'' in \emph{Proceedings of the 56th Annual Meeting of the Association for Computational Linguistics (Volume 1: Long Papers)}, 2018, pp. 2587--2597.

\bibitem{xu2019exact}
Y.~Xu, B.~Wu, F.~Shen, Y.~Fan, Y.~Zhang, H.~T. Shen, and W.~Liu, ``Exact adversarial attack to image captioning via structured output learning with latent variables,'' in \emph{Proceedings of the IEEE/CVF Conference on Computer Vision and Pattern Recognition}, 2019, pp. 4135--4144.

\bibitem{zhang2020fooled}
S.~Zhang, Z.~Wang, X.~Xu, X.~Guan, and Y.~Yang, ``Fooled by imagination: Adversarial attack to image captioning via perturbation in complex domain,'' in \emph{2020 IEEE International Conference on Multimedia and Expo (ICME)}.\hskip 1em plus 0.5em minus 0.4em\relax IEEE, 2020, pp. 1--6.

\bibitem{liObjectOrientedBackdoorAttack2022a}
M.~Li, N.~Zhong, X.~Zhang, Z.~Qian, and S.~Li, ``Object-oriented backdoor attack against image captioning,'' in \emph{ICASSP 2022 - 2022 IEEE International Conference on Acoustics, Speech and Signal Processing (ICASSP)}.\hskip 1em plus 0.5em minus 0.4em\relax Singapore, Singapore: IEEE, May 2022, pp. 2864--2868.

\bibitem{kwon2022toward}
H.~Kwon and S.~Lee, ``Toward backdoor attacks for image captioning model in deep neural networks,'' \emph{Security and Communication Networks}, vol. 2022.

\bibitem{vinyals2015show}
O.~Vinyals, A.~Toshev, S.~Bengio, and D.~Erhan, ``Show and tell: A neural image caption generator,'' in \emph{Proceedings of the IEEE conference on computer vision and pattern recognition}, 2015, pp. 3156--3164.

\bibitem{xu2015show}
K.~Xu, J.~Ba, R.~Kiros, K.~Cho, A.~Courville, R.~Salakhudinov, R.~Zemel, and Y.~Bengio, ``Show, attend and tell: Neural image caption generation with visual attention,'' in \emph{International conference on machine learning}.\hskip 1em plus 0.5em minus 0.4em\relax PMLR, 2015, pp. 2048--2057.

\bibitem{huang2019attention}
L.~Huang, W.~Wang, J.~Chen, and X.-Y. Wei, ``Attention on attention for image captioning,'' in \emph{Proceedings of the IEEE/CVF international conference on computer vision}, 2019, pp. 4634--4643.

\bibitem{vaswani2017attention}
A.~Vaswani, N.~Shazeer, N.~Parmar, J.~Uszkoreit, L.~Jones, A.~N. Gomez, {\L}.~Kaiser, and I.~Polosukhin, ``Attention is all you need,'' \emph{Advances in neural information processing systems}, vol.~30, 2017.

\bibitem{ren2015faster}
S.~Ren, K.~He, R.~Girshick, and J.~Sun, ``Faster r-cnn: Towards real-time object detection with region proposal networks,'' \emph{Advances in neural information processing systems}, vol.~28, 2015.

\bibitem{lin2017feature}
T.-Y. Lin, P.~Doll{\'a}r, R.~Girshick, K.~He, B.~Hariharan, and S.~Belongie, ``Feature pyramid networks for object detection,'' in \emph{Proceedings of the IEEE conference on computer vision and pattern recognition}, 2017, pp. 2117--2125.

\bibitem{liu2016ssd}
W.~Liu, D.~Anguelov, D.~Erhan, C.~Szegedy, S.~Reed, C.-Y. Fu, and A.~C. Berg, ``Ssd: Single shot multibox detector,'' in \emph{European Conference on Computer Vision}.\hskip 1em plus 0.5em minus 0.4em\relax Springer, 2016, pp. 21--37.

\bibitem{redmon2016you}
J.~Redmon, S.~Divvala, R.~Girshick, and A.~Farhadi, ``You only look once: Unified, real-time object detection,'' in \emph{Proceedings of the IEEE conference on computer vision and pattern recognition}, 2016, pp. 779--788.

\bibitem{thys2019fooling}
S.~Thys, W.~Van~Ranst, and T.~Goedem{\'e}, ``Fooling automated surveillance cameras: adversarial patches to attack person detection,'' in \emph{Proceedings of the IEEE/CVF conference on computer vision and pattern recognition workshops}, 2019, pp. 0--0.

\bibitem{li2021universal}
D.~Li, J.~Zhang, and K.~Huang, ``Universal adversarial perturbations against object detection,'' \emph{Pattern Recognition}, vol. 110, p. 107584, 2021.

\bibitem{si2023angelic}
W.~Si, S.~Li, S.~Park, I.~Lee, and O.~Bastani, ``Angelic patches for improving third-party object detector performance,'' in \emph{Proceedings of the IEEE/CVF Conference on Computer Vision and Pattern Recognition}, 2023, pp. 24\,638--24\,647.

\bibitem{gu2019badnets}
T.~Gu, K.~Liu, B.~Dolan-Gavitt, and S.~Garg, ``Badnets: Evaluating backdooring attacks on deep neural networks,'' \emph{IEEE Access}, vol.~7, pp. 47\,230--47\,244, 2019.

\bibitem{lyu2023poisoning}
X.~Lyu, Y.~Han, W.~Wang, J.~Liu, B.~Wang, J.~Liu, and X.~Zhang, ``Poisoning with cerberus: Stealthy and colluded backdoor attack against federated learning,'' in \emph{Proceedings of the AAAI Conference on Artificial Intelligence}, vol.~37, no.~7, 2023, pp. 9020--9028.

\bibitem{chan2022baddet}
S.-H. Chan, Y.~Dong, J.~Zhu, X.~Zhang, and J.~Zhou, ``Baddet: Backdoor attacks on object detection,'' in \emph{European Conference on Computer Vision}.\hskip 1em plus 0.5em minus 0.4em\relax Springer, 2022, pp. 396--412.

\bibitem{chen2021badnl}
X.~Chen, A.~Salem, D.~Chen, M.~Backes, S.~Ma, Q.~Shen, Z.~Wu, and Y.~Zhang, ``Badnl: Backdoor attacks against nlp models with semantic-preserving improvements,'' in \emph{Annual computer security applications conference}, 2021, pp. 554--569.

\bibitem{gan2021triggerless}
L.~Gan, J.~Li, T.~Zhang, X.~Li, Y.~Meng, F.~Wu, Y.~Yang, S.~Guo, and C.~Fan, ``Triggerless backdoor attack for nlp tasks with clean labels,'' \emph{arXiv preprint arXiv:2111.07970}, 2021.

\bibitem{qi2021turn}
F.~Qi, Y.~Yao, S.~Xu, Z.~Liu, and M.~Sun, ``Turn the combination lock: Learnable textual backdoor attacks via word substitution,'' \emph{arXiv preprint arXiv:2106.06361}, 2021.

\bibitem{qi2021hidden}
F.~Qi, M.~Li, Y.~Chen, Z.~Zhang, Z.~Liu, Y.~Wang, and M.~Sun, ``Hidden killer: Invisible textual backdoor attacks with syntactic trigger,'' \emph{arXiv preprint arXiv:2105.12400}, 2021.

\bibitem{pan2022hidden}
X.~Pan, M.~Zhang, B.~Sheng, J.~Zhu, and M.~Yang, ``Hidden trigger backdoor attack on $\{$NLP$\}$ models via linguistic style manipulation,'' in \emph{31st USENIX Security Symposium (USENIX Security 22)}, 2022, pp. 3611--3628.

\bibitem{huang2023training}
Y.~Huang, T.~Y. Zhuo, Q.~Xu, H.~Hu, X.~Yuan, and C.~Chen, ``Training-free lexical backdoor attacks on language models,'' in \emph{Proceedings of the ACM Web Conference 2023}, 2023, pp. 2198--2208.

\bibitem{tran2018spectral}
B.~Tran, J.~Li, and A.~Madry, ``Spectral signatures in backdoor attacks,'' \emph{Advances in neural information processing systems}, vol.~31, 2018.

\bibitem{gao2019strip}
Y.~Gao, C.~Xu, D.~Wang, S.~Chen, D.~C. Ranasinghe, and S.~Nepal, ``Strip: A defence against trojan attacks on deep neural networks,'' in \emph{Proceedings of the 35th Annual Computer Security Applications Conference}, 2019, pp. 113--125.

\bibitem{chen2018detecting}
B.~Chen, W.~Carvalho, N.~Baracaldo, H.~Ludwig, B.~Edwards, T.~Lee, I.~Molloy, and B.~Srivastava, ``Detecting backdoor attacks on deep neural networks by activation clustering,'' \emph{arXiv preprint arXiv:1811.03728}, 2018.

\bibitem{liu2018fine}
K.~Liu, B.~Dolan-Gavitt, and S.~Garg, ``Fine-pruning: Defending against backdooring attacks on deep neural networks,'' in \emph{International symposium on research in attacks, intrusions, and defenses}.\hskip 1em plus 0.5em minus 0.4em\relax Springer, 2018, pp. 273--294.

\bibitem{wu2021adversarial}
D.~Wu and Y.~Wang, ``Adversarial neuron pruning purifies backdoored deep models,'' \emph{Advances in Neural Information Processing Systems}, vol.~34, pp. 16\,913--16\,925, 2021.

\bibitem{zeng2021adversarial}
Y.~Zeng, S.~Chen, W.~Park, Z.~M. Mao, M.~Jin, and R.~Jia, ``Adversarial unlearning of backdoors via implicit hypergradient,'' \emph{{International Conference on Learning Representations (ICLR)}}, 2022.

\bibitem{zheng2022data}
R.~Zheng, R.~Tang, J.~Li, and L.~Liu, ``Data-free backdoor removal based on channel lipschitzness,'' in \emph{European Conference on Computer Vision}.\hskip 1em plus 0.5em minus 0.4em\relax Springer, 2022, pp. 175--191.

\bibitem{nguyen2021wanet}
A.~Nguyen and A.~Tran, ``Wanet--imperceptible warping-based backdoor attack,'' \emph{{International Conference on Learning Representations (ICLR)}}, 2021.

\bibitem{liu2020reflection}
Y.~Liu, X.~Ma, J.~Bailey, and F.~Lu, ``Reflection backdoor: A natural backdoor attack on deep neural networks,'' in \emph{European Conference on Computer Vision}.\hskip 1em plus 0.5em minus 0.4em\relax Springer, 2020, pp. 182--199.

\bibitem{zheng2020distance}
Z.~Zheng, P.~Wang, W.~Liu, J.~Li, R.~Ye, and D.~Ren, ``Distance-iou loss: Faster and better learning for bounding box regression,'' in \emph{Proceedings of the AAAI conference on artificial intelligence}, vol.~34, no.~07, 2020, pp. 12\,993--13\,000.

\bibitem{madry2017towards}
A.~Madry, A.~Makelov, L.~Schmidt, D.~Tsipras, and A.~Vladu, ``Towards deep learning models resistant to adversarial attacks,'' \emph{arXiv preprint arXiv:1706.06083}, 2017.

\bibitem{lin2014microsoft}
T.-Y. Lin, M.~Maire, S.~Belongie, J.~Hays, P.~Perona, D.~Ramanan, P.~Doll{\'a}r, and C.~L. Zitnick, ``Microsoft coco: Common objects in context,'' in \emph{European Conference on Computer Vision}.\hskip 1em plus 0.5em minus 0.4em\relax Springer, 2014, pp. 740--755.

\bibitem{zhou2020unified}
L.~Zhou, H.~Palangi, L.~Zhang, H.~Hu, J.~Corso, and J.~Gao, ``Unified vision-language pre-training for image captioning and vqa,'' in \emph{Proceedings of the AAAI conference on artificial intelligence}, vol.~34, no.~07, 2020, pp. 13\,041--13\,049.

\bibitem{papineni2002bleu}
K.~Papineni, S.~Roukos, T.~Ward, and W.-J. Zhu, ``Bleu: a method for automatic evaluation of machine translation,'' in \emph{Proceedings of the 40th annual meeting of the Association for Computational Linguistics}, 2002, pp. 311--318.

\bibitem{vedantam2015cider}
R.~Vedantam, C.~Lawrence~Zitnick, and D.~Parikh, ``Cider: Consensus-based image description evaluation,'' in \emph{Proceedings of the IEEE conference on computer vision and pattern recognition}, 2015, pp. 4566--4575.

\bibitem{lin2004rouge}
C.-Y. Lin, ``Rouge: A package for automatic evaluation of summaries,'' in \emph{Text summarization branches out}, 2004, pp. 74--81.

\bibitem{goodfellow2014explaining}
I.~J. Goodfellow, J.~Shlens, and C.~Szegedy, ``Explaining and harnessing adversarial examples,'' in \emph{International Conference on Learning Representations}, 2015.

\bibitem{carlini2017towards}
N.~Carlini and D.~Wagner, ``Towards evaluating the robustness of neural networks,'' in \emph{2017 ieee symposium on security and privacy (sp)}.\hskip 1em plus 0.5em minus 0.4em\relax Ieee, 2017, pp. 39--57.

\bibitem{selvaraju2017grad}
R.~R. Selvaraju, M.~Cogswell, A.~Das, R.~Vedantam, D.~Parikh, and D.~Batra, ``Grad-cam: Visual explanations from deep networks via gradient-based localization,'' in \emph{Proceedings of the IEEE international conference on computer vision}, 2017, pp. 618--626.

\bibitem{qi2021onion}
F.~Qi, Y.~Chen, M.~Li, Y.~Yao, Z.~Liu, and M.~Sun, ``Onion: A simple and effective defense against textual backdoor attacks,'' in \emph{Proceedings of the 2021 Conference on Empirical Methods in Natural Language Processing}, 2021, pp. 9558--9566.

\end{thebibliography}

\vspace{5pt}

\vfill

\end{document}